\documentclass[12pt]{JHEP3}
\usepackage{amsmath,epsfig,graphicx}

\newcommand{\be}{\begin{equation}}
\newcommand{\ee}{\end{equation}}
\newcommand{\bq}{\begin{eqnarray}}
\newcommand{\eq}{\end{eqnarray}}
\newcommand{\bsq}{\begin{subequations}}
\newcommand{\esq}{\end{subequations}}
\newcommand{\bc}{\begin{center}}
\newcommand{\ec}{\end{center}}

\newcommand{\pd}{\partial}
\renewcommand{\thesection}{\arabic{section}}

\title{Non-Relativistic Strings in Expanding Spacetime}

\author{A. Avgoustidis, Joaquim Gomis \\
{\it Departament d'Estructura i Costituents de la Mat\`eria,\\
Facultat de F\'isica, Universitat de Barcelona,\\
Diagonal 647, 08028 Barcelona, Spain\\}
\email{tasos, gomis@ecm.ub.es}}

\abstract{
We obtain a non-relativistic diffeomorphism invariant string action
as a special limit of the Nambu-Goto action in an FLRW background.  
We use this action to study non-relativistic string dynamics in an 
expanding universe and construct an analytic model describing the 
macroscopic properties of non-relativistic string networks.  The 
non-relativistic constraint equations allow arbitrarily small string 
velocities and thus a `frustrated' equation of state for non-interacting 
strings can be obtained without the need of a velocity damping
mechanism.  Assuming that colliding string segments reconnect by 
exchange of partners, non-relativistic string networks exhibit 
scaling behaviour, but with enhanced energy densities due to the 
smaller average string velocity.  Non-relativistic string networks 
can be relevant in several contexts in condensed matter physics and 
cosmology.}

\preprint{UB-ECM-PF-07-19}

\begin{document}

 \section{\label{intro}Introduction}

  Progress in solving string theory in various backgrounds can be 
  done by considering  sectors of the theory that decouple from 
  the rest of the degrees of freedom in a suitable limit. Such 
  decoupled sectors are characterized by having an altogether 
  different asymptotic symmetry compared to that  of the parent 
  string theory.  A well known example to such a truncation is 
  the BMN \cite{Berenstein:2002jq} sector of  string theory in 
  AdS$_5\times S^5$.  Once a consistent sector is found, a 
  complete worldsheet theory  with the appropriate symmetries can  
  be written down without further reference of the parent theory.

  Non-relativistic string theory  \cite{jGom_Oog} (see also
  \cite{DanGuiKru})   in flat space is another consistent
  sector of string theory, whose world-sheet conformal field 
  theory description has the appropriate Galilean symmetry
  \cite{Brugues:2004an}. Non-relativistic superstrings and
  non-relativistic superbranes \cite{JGom_Kam_Town,Garcia:2002fa} 
  are obtained as a certain decoupling limit of the full relativistic
  theory. The basic idea behind the decoupling limit is to take a
  particular non-relativistic limit in such a way that the light
  states satisfy a Galilean-invariant dispersion relation, while the
  rest decouple. For the case of strings, this can be accomplished by
  considering wound strings in the presence of a background $B$-field
  and tuning the $B$-field so that the energy coming from the
  $B$-field cancels the tension of the string. In flat space,  
  once kappa symmetry and diffeomorphism invariance are fixed,
  non-relativistic strings are described by a free field theory 
  in flat space. In AdS$_5\times S^5$ \cite{GGK_AdS5XS5}, the
  world-sheet theory reduces to a supersymmetric free field theory 
  in AdS$_2$.

  It is an interesting question whether similar non-relativistic
  string actions can be constructed in an expanding spacetime
  and if so, whether non-relativistic strings could play a
  cosmological role in the form of cosmic strings.  The study 
  of cosmic strings\footnote{The dynamics of cosmic strings can be
  described by considering perturbations around a static solitonic
  string solution.  Keeping all orders in the perturbations results
  in a relativistic effective string action, while keeping only up
  to quadratic order gives rise to the non-relativistic string action
  we will consider in this paper \cite{JGom_Kam_Town}.} has been  
  catalysed in the past few years mainly due to theoretical 
  motivations, in particular 
  the realisation that they are generic in Supersymmetric Grand 
  Unified Theory (SUSY GUT) models \cite{Jeannerot} and brane 
  inflation \cite{BMNQRZ,SarTye}.  The latter possibility is of 
  particular significance as it provides a potential observational 
  window to superstring physics \cite{PolchStab,PolchIntro}.  Further, 
  the fact that the Planck satellite and laser interferometers such 
  as LISA and LIGO may be able to probe a significant part of cosmic
  string tensions relevant to these models \cite{DamVil}, opens
  the possibility of detecting cosmic strings in the foreseeable
  future.

  One can think of situations in which ordinary cosmic strings
  could behave non-relativistically.  Network simulations in
  a matter or radiation dominated universe \cite{All_Shell} 
  suggest that, at late times, string segments move relatively 
  slowly and coherently on the largest scales, but also show 
  evidence that small-scale-structure \cite{Mart_Shell_sss,Polch_Rocha}
  which is largely responsible for damping energy from the network
  through the formation of minuscule loops, remains relativistic
  as Hubble damping is inefficient at scales much smaller than the 
  horizon \cite{book}.  However, the situation is different for 
  strings in de Sitter spacetime, where Hubble damping can be very 
  efficient rendering the strings essentially non-relativistic.  
  This may be relevant for late time cosmology as observations 
  \cite{Perlmutter,WMAP3} suggest that the universe is already 
  entering a de Sitter phase.  Further, non-relativistic string  
  networks have been considered as Solid Dark Matter (SDM)  
  \cite{BuchSper,BatCarChMo} and more recently \cite{Alexand}  
  as an alternative explanation of galactic rotation curves.

  It would thus be desirable to have an effective diffeomorphism
  invariant action\footnote{Note that Ref.~\cite{BuchSper} considers
  an action applicable to a `continuous medium' with internal
  structure, which is invariant under limited reparametrisations
  preserving the worldlines of the constituent particles.  Here
  we consider a \emph{diffeomorphism invariant} non-relativistic
  action.} describing the dynamics of non-relativistic strings
  in a cosmological context.  On the other hand, the fact that one
  can construct a consistent worldsheet theory of non-relativistic
  strings at quantum level (in flat space) also motivates the
  study of \emph{fundamental} non-relativistic strings in an
  expanding spacetime.  In this paper, we point out that a
  non-relativistic diffeomorphism invariant action can be obtained
  in the case of a Friedmann-Lema\^{i}tre-Robertson-Walker (FLRW)
  spacetime as a limit of the relativistic Nambu-Goto action, and
  study the dynamics and cosmological evolution of non-relativistic 
  strings.

  The structure of the paper is as follows.  In section
  \ref{rel_string} we review some basic results about the
  Nambu-Goto action and the dynamics of relativistic strings
  in an expanding universe, which will be useful for comparison
  to the non-relativistic case.  In section \ref{non_rel_lim} we
  obtain a non-relativistic diffeomorphism invariant worldsheet
  action by taking a particular limit of the Nambu-Goto action
  in expanding spacetime.  We move on in section \ref{NR_dynamics}
  to study non-relativistic string dynamics as described by this
  action.  The physical interpretation of non-relativistic strings
  as well as their possible coupling to cosmology-in particular
  the effective equation of state of an `ideal gas' of
  non-interacting, non-relativistic strings-are discussed in
  section \ref{physical}.  The effect of string interactions is
  left for section \ref{VOS}, where macroscopic models for the
  cosmological evolution of both relativistic and non-relativistic 
  string networks are discussed.  In section \ref{RelVnonRel} we 
  solve numerically the non-relativistic network model for a wide 
  range of parameters and compare the results to those of the  
  relativistic model.  In section \ref{discuss} we discuss possible  
  applications of non-relativistic strings in condensed matter and 
  cosmological contexts.  Finally, we have three appendices which 
  describe an alternative derivation of our non-relativistic string 
  action as a semiclassical expansion \cite{semiclass} around the 
  vacuum solution (appendix \ref{semicl}), the Hamiltonian formulation  
  of relativistic and non-relativistic string dynamics (appendix  
  \ref{hamiltonian}) and the construction of a spacetime energy-momentum  
  tensor for the non-relativistic string (appendix \ref{NR_EM_tensor}).

 \section{\label{rel_string} Relativistic String in Expanding Spacetime}\setcounter{equation}{0}

 Let us first consider a string moving in a $D+1$ dimensional
 spacetime with metric $G_{MN}\, \,(M,N=0,1,2,\ldots,D)$.  Its
 world history is described by a two-dimensional spacetime surface,
 the string worldsheet $x^M=x^M(\sigma^i)$, $i=0,1$.
 The dynamics is governed by the Nambu-Goto action
 \be\label{nambu}
  S_{NG}=-T_R \! \int \! \sqrt{-\gamma}\, d^2\sigma \, ,
 \ee
 where $T_R$ is the string tension and $\gamma$ is the determinant
 of the pullback of the background metric on the worldsheet,
 $\gamma_{ij}=G_{MN}(x)\pd_i x^M \pd_j x^N$.

 The equations of motion for the fields $x^M$ obtained from this
 action are given by
 \be\label{eom}
  \nabla^2 x^M + \Gamma^M_{N\Lambda} \gamma^{ij}
  \pd_i x^N \pd_j x^\Lambda = 0 \, ,
 \ee
 where $\Gamma^M_{N\Lambda}$ is the ($D+1$)-dimensional Christoffel
 symbol and $\nabla^2 x^M$ the covariant Laplacian of the worldsheet
 fields $x^M$.

 By varying the action with respect to the background metric $G_{MN}$
 we obtain a spacetime energy-momentum tensor
 \be\label{emt}
  T^{MN}(y^\Lambda)=\frac{1}{\sqrt{-G}}\,T_R \! \int \! d^2\sigma
  \sqrt{-\gamma} \gamma^{ij}\pd_i x^M
  \pd_j x^N \, \delta^{(D+1)}(y^\Lambda-x^\Lambda(\sigma^i))\,.
 \ee
 The rigid symmetries of (\ref{nambu}) are given by the Killing
 vectors of $G_{MN}$.  The Nambu-Goto action is also invariant
 under 2D diffeomosrphisms of the worldsheet coordinates $\sigma^i$.
 One can use this freedom to fix the gauge by imposing two conditions
 on $x^M(\sigma^i)$.

 Now consider string propagation in an expanding Universe
 described by a flat FLRW metric
 \be\label{FLRW_metric}
  G_{MN} = a(x^0)^2 \eta_{MN}
 \ee
 in conformal time $x^0\!\equiv\!\eta$.

 A convenient gauge choice in this case is the \emph{transverse temporal
 gauge} given by:
 \be\label{trans_temp}
  \left\{
   \begin{array}{l}
    \dot x x^\prime = 0 \\
    \tau=x^0
   \end{array}
  \right.
 \ee
 The equations of motion (\ref{eom}) become~\cite{Tur_Bhatt}:
 \bq
  \left\{
   \begin{array}{l}
    \dot\epsilon=-2\frac{\dot a}{a} \epsilon \dot{\bf x}^2
     \label{eom_eps} \\
    \ddot {\bf x} + 2\frac{\dot a}{a} \left( 1 - \dot{\bf x}^2 \right)
     \dot {\bf x} = {\left(\frac{{\bf x}^{\prime}}{\epsilon}\right)}^{\prime}
     \epsilon^{-1} \label{eom_x} 
   \end{array}
   \right.
  \eq
  where
  \be\label{epsilon}
   \epsilon = \frac{ {-x^\prime}^2 }{ \sqrt{-\gamma} } =  \left(
   \frac{ {{\bf x}^\prime }^2 } { 1-\dot{\bf x}^2 } \right)^{1/2}\, .
  \ee

  The variable $\epsilon$ is related to the canonical momentum
  associated to the field $x^0(\tau)$.  Indeed, in the transverse
  gauge $\gamma_{01}\!\equiv\!\gamma_{\tau\sigma}\!=\!0$, we have:
  \be\label{p_0_Lagr}
   p_0=-\frac{T_R}{2} \sqrt{-\gamma} \gamma^{ij} \frac{\pd \gamma_{ij}}
   {\pd_\tau x^0}= - T_R a(x^0)^2 \epsilon \dot x^0 \, ,
  \ee
  which, after imposing the temporal gauge condition $\dot x^0=1$,
  becomes $-T_R a(x^0)^2 \epsilon$.

  In the transverse temporal gauge, the energy-momentum tensor (\ref{emt})
  of a relativistic string in a FLRW background is \cite{book}:
  \be\label{emt_trans_temp}
   T^{MN}(\eta,y^I)=\frac{1}{a(\eta)^{D+1}}\,T_R \! \int \! d\sigma
  \left( \epsilon \dot x^M \dot x^N -\epsilon^{-1} x^{\prime M}
  x^{\prime N} \right) \, \delta^{(D)}(y^I-x^I(\sigma,\eta))\, ,
  \ee
  where, having integrated out $\delta(\eta-x^0(\tau))$, $I$ runs from
  $1$ to $D$.

  To construct the string energy one projects $T^{MN}$ on a spatial
  hypersurface $\eta\!=\!{\rm const}$, with induced metric $h$ and normal
  covectors $n_M\!=\!\left(a(\eta),{\bf 0}\right)$, integrating over the
  $D$ spatial coordinates:
  \bq\label{energy}
   E(\eta) &=& - \! \int \! \sqrt{h} n_M n_N T^{MN} d^D{\bf y} \nonumber \\
          &=& \! \int \! \sqrt{h} T^0_{\; 0} d^D{\bf y} \, .
  \eq
  Thus, due to the foliation, the energy can be constructed from the
  $00$ component of the energy momentum tensor. Since, $\sqrt{h}=
  a(\eta)^D$, equation (\ref{energy}) becomes:
  \be\label{string_energy}
   E(\eta) = \! \int \! T^0_{\;0} a(\eta)^D d^D{\bf y}
   = a(\eta) T_R \! \int \! \epsilon \, d\sigma \, .
  \ee
  Therefore, the energy of the relativistic string is simply the
  tension times the {\it physical} string length, taking into account
  relativistic length contraction.

 \section{\label{non_rel_lim}Non-Relativistic Limit of Nambu-Goto
          Action in FLRW}\setcounter{equation}{0}

 Now consider a string, charged under a background antisymmetric
 2-tensor field $B_{MN}$, propagating in $D+1$ FLRW spacetime.
 The string couples to $B$ through a topological Wess-Zumino term,
 so that the total action reads:
 \be\label{total_action}
  S=S_{NG}+S_{WZ}=-T_R \! \int \! \sqrt{-\gamma}\, d^2\sigma
    + q \! \int \! B^* \, ,
 \ee
 where $q$ is the string charge and $B^*$ the pullback of $B$
 on the worldsheet.  We consider a relativistic string aligned in
 the $x^0$, $x^1$ directions, its transverse coordinates being
 $X^a$ with $a=2,3,...D$.

 The non-relativistic limit \cite{jGom_Oog, JGom_Kam_Town, DanGuiKru}
 of this string consists of rescaling the longitudinal coordinates
 \be\label{rescale}
  x^\mu \rightarrow \omega x^\mu\,, \quad \mu=0,1
 \ee
 and taking the limit $\omega \rightarrow \infty$.  This yields a
 divergent term, coming from $S_{NG}$, which (in some geometries)
 can be canceled by an appropriate choice of a closed $B_{MN}$.
 If we assume that the string is wrapped on a spatial circle
 \be\label{circle}
  x^1 \sim x^1 + 2\pi R
 \ee
 then the chosen $B_{01}$ cannot be set to zero by a gauge
 transformation.

 The above procedure generally works in flat spacetime (in fact
 one only needs the longitudinal part of the metric be flat
 \cite{JGom_Kam_Town}), but for a curved background it is not
 guaranteed that there is a choice of closed $B$ which cancels
 the diverging piece of the action.  Non-relativistic
 superstring actions have been obtained in the case of
 AdS$_5\times S^5$ \cite{GGK_AdS5XS5}.

 We will now see that the non-relativistic limit can also be taken
 in the case of a FLRW background.  We write the Lagrangian density
 of the Nambu-Goto piece as:
 \bq 
  {\cal{L}}_{NG} &=& -T_R \sqrt{-{\rm det}[g_{ij}+G_{ab}(\eta)\pd_i X^a
  \pd_j X^b]} \nonumber \\
  &=&-T_R \sqrt{-{\rm det}g_{ij}}\sqrt{{\rm det}[\delta^i_j+g^{ik}
  G_{ab}(\eta)\pd_k X^a \pd_j X^b]} \label{L_NG} \, ,
 \eq 
 where $g_{ij}=G_{\mu\nu}\pd_i x^\mu \pd_j x^\nu$ and $G_{ab}(\eta)=
 a(\eta)^2 \delta_{ab}$.

 Then, assuming a power law expansion $a(\eta)\!=\!\eta^{\alpha/2}$
 (for example $\alpha=2$ resp. $4$ in radiation resp. matter dominated
 era) we obtain the non-relativistic limit of $S_{NG}$ by the rescaling
 (\ref{rescale}), which implies
 \be
  a(\eta) \rightarrow \omega^{\alpha/2} a(\eta)\label{a_rescale} \, .
 \ee

 Expanding the Lagrangian density in powers of the parameter $\omega$
 we then obtain:
 \be\label{L_NG_expand}
  {\cal{L}}_{NG} = -T_R \omega^\alpha \left\{ \omega^2 \sqrt{-{\rm det}g}
  + \frac{1}{2}\sqrt{-{\rm det}g} g^{ij}G_{ab}(\eta)\pd_i X^a
  \pd_j X^b + {\cal O}\left(\frac{1}{\omega^2}\right)\right\} \, .
 \ee
 We can then rescale the string tension by
 \be\label{T_rescale} 
  T_R\, \omega^\alpha \rightarrow T_0
 \ee 
 and take the limit $\omega \rightarrow \infty$, yielding a finite
 and a divergent piece:
 \bq 
  {\cal L}_{\rm reg}&=&-\frac{T_0}{2} \sqrt{-{\rm det}g} g^{ij}
  G_{ab}(\eta)\pd_i X^a \pd_j X^b \label{L_reg} \\
  {\cal L}_{\rm div}&=&-T_0\omega^2\sqrt{-{\rm det}g} =-T_0\omega^2
  a(\eta)^2 \sqrt{-{\rm det}(\eta_{\mu\nu}\pd_i x^\mu
  \pd_j x^\nu)} \label{L_div} \, .
 \eq 
 The divergent piece can be canceled by choosing an appropriate closed
 $B_{\mu\nu}$.  Indeed if we choose\footnote{The chosen $B_{\mu\nu}$
 is closed.  Working in zweibeins $e^\mu$ we have: ${\rm d}B =
 \frac{1}{2}{\rm d}[a(x^0)^2\epsilon_{\mu\nu}e^\mu \wedge
 e^\nu] = {\rm d} [ a(x^0)^2 e^0 \wedge e^1 ] = 2a \dot a \, e^0
 \wedge e^0 \wedge e^1 + a^2 ({\rm d}e^0 \wedge e^1 + e^0 \wedge
 {\rm d}e^1 ) = a^2 (-w^{01} \wedge e^1 \wedge e^1 ) - a^2 e^0
 \wedge w^{10} \wedge e^0 = 0$ where we have used ${\rm d}e + 
 w\wedge e=0$, Cartan's structure equation with zero torsion.} 
 $B_{\mu\nu}=a(\eta)^2 \epsilon_{\mu\nu}$ the Wess-Zumino part 
 of the Lagrangian becomes:
 \be\label{L_WZ}
  \frac{1}{2} \omega^2 (q\omega^\alpha) a(\eta)^2 \epsilon_{\mu\nu}
  \epsilon^{ij} \pd_i x^\mu \pd_j x^\nu .
 \ee
 This term precisely cancels the divergent piece (\ref{L_div}) if one
 tunes the rescaled charge $(q\omega^\alpha)$ with the string tension
 $T_0$.  We are thus left with the Non-Relativistic string action:
 \be\label{S_NR}
  S_{NR}=-\frac{T_0}{2} \! \int \! \sqrt{-{\rm det}g}
  g^{ij}G_{ab}(\eta)\pd_i X^a \pd_j X^b d^2\sigma \, .
 \ee
 This action can also be derived by a `semiclassical approximation'
 \cite{semiclass}
 from the classical solution:
 \be
  x_0^M = \left\{
   \begin{array}{cll}
           \tau       &,&      M=0           \\
       \lambda\sigma  &,&      M=1           \\
             0        &,&  M=a\in (2,\ldots,D)
   \end{array}
  \right.
 \ee
 (see Appendix~\ref{semicl} for details).

 \section{\label{NR_dynamics}Non-Relativistic String Dynamics}\setcounter{equation}{0}

 The action (\ref{S_NR}) is characterised by 2D diffeomorphism invariance
 with respect to the worldsheet coordinates $\sigma^i$ and global Galilei
 invariance (modulo time translations due to time dependence of the
 metric) with respect to the transverse spacetime coordinates $X^a$.
 The canonical variables satisfy two primary constraints:
 \bq
  && p_\mu \epsilon^{\mu\rho} \eta_{\rho\nu} x^{\prime\nu} + \frac{1}{2}
  \left(\frac{P_aP_b}{T_0}G^{ab}(x^0) + T_0 X^{\prime a}X^{\prime b}
  G_{ab}(x^0) \right) = 0  \label{energy_constr}  \\
  && p_\mu x^{\prime\mu} + P_a X^{\prime a} = 0 \, ,  \label{pxp_constr}
 \eq
 where $p_\mu$, $P_a$ are the canonical momenta corresponding to $x^\mu$
 and $X^a$.

 Varying the action with respect to the transverse and longitudinal fields
 $X^a$ and $x^\mu$ respectively, one obtains the equations of motion:
 \bq
  \begin{array}{l}
   \pd_i\left(\sqrt{-g}g^{ij}\pd_j X^a\right)+\sqrt{-g}g^{ij}
   \Gamma^a_{bc} \pd_i X^b \pd_j X^c + \sqrt{-g}g^{ij}
   \pd_i X^b \pd_j x^0 \, (\pd_0 G_{bc}) \, G^{ca} = 0
  \end{array}
  \nonumber \\ 
  && \label{eom_trans} \\
  \begin{array}{rl}
    \pd_i\left[\sqrt{-g}\pd_kx^\nu\eta_{\mu\nu} a(x^0)^2 \left(
    g^{ik}g^{mn}-2g^{im}g^{kn} \right) \right. \pd_m & \left. \!\! 
    X^a \, \pd_n X^b \, G_{ab}(x^0) \right]  \\
    & = \sqrt{-g} g^{mn} \pd_m X^a \pd_n X^b \frac{\pd G_{ab}(x^0)} 
    {\pd x^\mu} 
  \end{array}
  \nonumber \\
 && \label{eom_long} 
 \eq
 subject to the boundary condition (\ref{circle}).
 For the metric (\ref{FLRW_metric}) the Christoffel symbols
 $\Gamma^a_{bc}$ vanish and the transverse equations of motion
 (\ref{eom_trans}) relate the covariant divergence of the
 transverse fields $X^a$ to the time derivative of the transverse
 metric $G_{ab}$.  We can use the 2D reparametrisation invariance
 of the action to fix the gauge.  For our discussion it will be
 convenient to work in the \emph{static gauge}
 \be\label{nonrel_gauge}
  x^0-\tau=0 \, , \quad x^1-\lambda\sigma=0 \, ,
 \ee
 identifying worldsheet and background times, while allowing for
 multiple windings of the non-relativistic string.  Indeed, defining
 $\sigma\in [0,2\pi)$, the periodicity condition (\ref{nonrel_gauge})
 requires that
 \be\label{lambda}
  \lambda=nR \, ,
 \ee
 where $n$ is the string winding number.

 After fixing the gauge, the physical degrees of freedom of the
 non-relativistic string are the transverse coordinates $X^a$
 and the corresponding momenta $P_a$.  The equation of motion
 (\ref{eom_trans}) becomes:
 \be\label{eom_gauge}
  \ddot X^a = -2\frac{\dot a}{a} \dot X^a + \lambda^{-2}
  X^{\prime\prime a} \, ,
 \ee
 which is the wave equation with a cosmological damping term
 $-2\frac{\dot a}{a} \dot X^a$.  This equation (for $\lambda\!=\!1$)
 has been used by Vilenkin \cite{Vil_CS} to describe small
 perturbations around a straight cosmic string, and was obtained
 by taking the limit $\dot X^2\!\ll\! 1$, $X^{\prime 2}\!\ll\! 1$
 of the relativistic equations of motion in the static gauge.
 Here, there is also a winding number $\lambda$.  One might be
 tempted to say that, for $\dot a/a=0$, equation (\ref{eom_gauge})
 implies a wave propagation velocity
 \be\label{v_0}
  v_0^2=\lambda^{-2} \, .
 \ee
 However, one should remember that the physical coordinates are not
 $\sigma,\tau$ but rather $x^1\!=\!\lambda \sigma,\, x^0=\tau$, so
 rewriting (\ref{eom_gauge}) in terms of the physical variables
 we get (in the case $\dot a/a=0$)
 \be\label{eom_phys}
  \pd_{x^0}^2 X^a = \pd_{x^1}^2 X^a \, ,
 \ee
 which describes a wave propagating at the velocity of light.  The
 non-relativistic string allows the propagation of waves along the
 longitudinal directions with the speed of light.  However, the
 transverse velocities are not restricted, in contrast to the case
 of the relativistic string.

 An `energy'
 \be\label{P_0}
  {\cal P}_0=\frac{1}{2\lambda} \! \int \! d\sigma \left(\frac{P_aP_b}{T_0}
  G^{ab}(x^0) + T_0 X^{\prime a}X^{\prime b} G_{ab}(x^0) \right)
 \ee
 can be obtained from the constraint (\ref{energy_constr}), which
 in the gauge (\ref{nonrel_gauge}) becomes:
 \be\label{P_0_gauge}
  {\cal P}_0 = \frac{1}{2} \! \int \! d\sigma \left( \lambda T_0
  \dot X^a \dot X^b + \lambda^{-1} T_0 X^{\prime a}X^{\prime b}
  \right) G_{ab}(x^0) \, .
 \ee
 This can be interpreted as the sum of the kinetic and potential
 energies of transverse excitations along the string.  The actual
 string energy, obtained by integrating the projection of the
 energy-momentum tensor on a constant $x^0$ hypersurface is (see
 appendix \ref{NR_EM_tensor}):
 \be\label{E_nonrel}
  E(x^0)=a(x^0)^{-1}\,{\cal P}_0 \, .
 \ee

 Since $x^0$-translation is not an isometry of (\ref{FLRW_metric})
 the Lagrangian is not time-translationally invariant and $p_0$ is not
 conserved.  In fact, its time evolution can be found from the
 longitudinal equations of motion (\ref{eom_long}).
 In the gauge (\ref{nonrel_gauge}) the $\mu=0$ component of
 (\ref{eom_long}) becomes:
 \be\label{p_0_dot}
  \frac{1}{2} \left(\dot X^a \dot X_a + \lambda^{-2} X^{\prime a}
  X^{\prime}_a \right)\dot{} = \lambda^{-2}\left(X^{\prime a}
  \dot X_a\right)^\prime + \frac{\dot a}{a} \left( \lambda^{-2}
  X^{ \prime a} X^{\prime}_a - \dot X^a \dot X_a \right) \, .
 \ee
 Integrating we obtain:
 \be\label{P_0_dot}
  \dot{\cal P}_0 = a \dot a \lambda T_0 \!\int\! d\sigma \left(
  \lambda^{-2} X^{ \prime a} X^{\prime b} - \dot X^a \dot X^b
  \right)\delta_{ab} \, ,
 \ee
 where the boundary term gives no contribution due to the
 periodicity condition (\ref{circle}).

 Similarly, from the constraint (\ref{pxp_constr}) we define the
 momentum ${\cal P}_1$ along the string
 \be\label{P_1}
  {\cal P}_1=-\frac{1}{\lambda} \! \int \! P_a X^{\prime a} d
  \sigma = - T_0 \! \int \! \dot X_a X^{\prime a} d\sigma
 \ee
 in the gauge (\ref{nonrel_gauge}).  Translational invariance
 then dictates that ${\cal P}_1$ is conserved, as can be easily
 verified using the equations of motion.

 \section{\label{physical}Physical Interpretation and Cosmology}\setcounter{equation}{0}

 \subsection{The NR Particle vs NR String Limit}

 The non-relativistic limit is generally understood as a low velocity
 limit, which can be formally obtained by sending the speed of light
 $c$ to infinity.  This procedure works, at least in the case of the
 point particle, although there are some conceptual issues involved
 when taking limits of dimensionful constants like $c$
 \cite{Duff,DufOkVen}.  A safer route is to keep $c$ constant and
 rescale the time coordinate by a dimensionless parameter, say $\omega$,
 taking the limit $\omega\rightarrow\infty$.  One can thus obtain a
 reparametrisation invariant, non-relativistic action for the point
 particle.  The naive application of this to the case of the string
 fails\footnote{The string obtained in this limit has a fixed length
 and no physical oscillations (see Ref.~\cite{Yastremiz}).} but this
 problem was solved with the realisation \cite{jGom_Oog} (see also 
 \cite{DanGuiKru}) that in order to obtain a Galilei invariant string  
 action one has to rescale both longitudinal coordinates, not the time  
 coordinate only.  In a sense, one can speak of a non-relativistic  
 `particle' limit, obtained by taking $v\ll 1$ and a non-relativistic  
 limit for extended objects for which one has to rescale all worldvolume  
 coordinates, as we did in section \ref{non_rel_lim} for the case of 
 the string.

 The rescaling of the longitudinal string direction corresponds
 to the assumption $(\pd y/ \pd x)^2\ll 1$, which one makes when
 deriving the wave equation by applying Newton's 2nd law on an
 infinitesimal string segment.  In the rescaling prescription
 we followed, the waves move along the string at the speed of
 light as the string tension equals the mass per unit length.  
 One usually thinks of non-relativistic strings as `violin-type' 
 having a small tension compared to their mass per unit length  
 and thus a subluminal `sound speed' along the string.  In this 
 sense, the strings we consider here are `hybrid', having a 
 relativistic speed of propagation along the string, but transverse
 Galilei invariance.  However, it is precisely this hybrid action
 (in flat space) which arises in the simplest Lorentz invariant field 
 theories when one studies the low-energy dynamics of domain wall 
 solutions.  Strings with subluminal propagation speeds (which would
 correspond to a differentiation between the string mass per unit 
 length and the tension) can arise in more complicated models, which 
 allow for spontaneously broken longitudinal Lorentz invariance 
 through a current generation mechanism on the string 
 worldsheet\footnote{See Ref.~\cite{BlPil_Redi} for a discussion
 of the relation between strings with broken longitudinal 
 Lorentz invariance and Kaluza-Klein strings in one dimension 
 higher.} \cite{Witten, Carter_cwc}.  To obtain such string actions 
 as a non-relativistic limit of the Nambu-Goto action, one would
 have to rescale each of the longitudinal coordinates by a  
 different factor and take both factors to infinity while keeping 
 their ratio constant.

 Note that, in order to ensure that the antisymmetric field $B$ used
 to cancel the divergent piece of the action cannot be gauged away,
 the non-relativistic string had to wind a compact dimension, say
 $x^1\sim x^1 + 2\pi R$.  In fact, the divergent piece of the action
 is a total derivative with respect to the worldsheet coordinates
 \cite{JGom_Kam_Town}, so if the action (\ref{S_NR}) was to be
 interpreted as an effective non-relativistic action, one could
 simply drop this term without requiring that the string is wound.
 However, if the non-relativistic string is to be interpreted as a 
 fundamental object, consistency requires a non-trivial winding.  In  
 this case, there are two distinct scales, namely $T_0$ of dimension
 mass-squared, which appears in the action (\ref{S_NR}), and the mass
 scale $m=2\pi n R T_0$, related to the geometry (through the
 compactification radius $R$) and the string winding number $n$.
 In fact, when quantising the non-relativistic string \cite{jGom_Oog},
 one encounters again the necessity of winding, as the mass $m$ is
 needed to define the energy states of the non-relativistic string
 spectrum.  In the flat case there are no physical states with
 zero winding number \cite{jGom_Oog}.
 Also note that in deriving the non-relativistic string action
 (\ref{S_NR}) we have defined the tension $T_0$ by a rescaling
 of the relativistic string tension $T_R$, appearing in the
 Nambu-Goto action (see equation (\ref{T_rescale})):
 \be\label{T0_TR}
  T_0 = \omega^\alpha T_R \, ,
 \ee
 where the expansion exponent $\alpha$ is positive and $\omega$ is
 taken to infinity.  Interpreting $T_0$ as the physically relevant
 quantity which is to be kept constant, the relativistic tension
 $T_R$ goes to zero as $\omega$ tends to infinity. 

 Finally, we comment on the stability of the non-relativistic string.
 A closed non-relativistic string is more stable to breakage than
 its relativistic counterpart\footnote{We thank F. Passerini for
 discussions of this point.}.  This is a consequence of the winding,
 which only allows a discrete number of potential `splitting points'
 along the string.  From an astrophysical perspective, ordinary
 cosmic string loops decay through gravitational radiation, which
 mainly couples to the kinetic energy of the fluctuations.  In
 particular, the power in gravitational radiation scales with the
 sixth power of the root mean square (rms) velocity (see for example
 \cite{book, vosk}).  Thus, if such non-relativistic strings were
 to play an astrophysical role, their decay rate would be power-law
 suppressed.  For long stings, the main energy-loss mechanism is
 through string intercommutation, which removes string length from
 the long string network.  This is also expected to be suppressed
 for non-relativistic strings as the interaction rates are
 proportional to the string velocities.  We shall now consider
 the possibility of coupling non-relativistic strings to cosmology.

 \subsection{Coupling to Gravity and Cosmology}

 The non-relativistic action we have analysed describes the dynamics
 of the independent degrees of freedom of the non-relativistic string,
 namely the transverse excitations.  In obtaining this action we have
 introduced a closed $B$ field, which cancels the divergent piece
 corresponding to the rest energy of the string.  Alternatively, if
 we are not interested in quantisation, we can simply drop the
 divergent part of the action-without introducing the $B$ field-because
 it is a total derivative (cf the case of the point-particle).
 Here, we will follow the latter approach.  The energy-momentum
 tensor of the non-relativistic string (Appendix \ref{NR_EM_tensor})
 therefore describes the energy of the transverse excitations but does
 not include a contribution from the rest mass of the string.  However,
 when one couples non-relativistic matter to General Relativity
 it is necessary to include the rest mass $m_0 c^2$ in the
 energy-momentum tensor, which gives the main contribution to
 the $T^{00}$ part while kinetic contributions are subdominant.
 Following this logic we will add the rest mass of the string to
 the $T^{00}$ part of the energy-momentum tensor of Appendix
 \ref{NR_EM_tensor}, which can then be coupled to Einstein's
 equations.  From now on we work in $D=3$ spatial dimensions. 

 Consider a cosmological setup where the cosmic fluid has a
 component due to a gas of non-interacting, non-relativistic strings.
 To obtain the energy density of the string fluid, one has to sum the
 contributions of all string segments in the network and, as we
 discussed, it is the rest energy of the segments which will give the
 dominant contribution.  This is in analogy to a gas of non-relativistic
 particles (dust), where the dominant contribution to the energy
 density is $\rho \equiv T^0_{\; 0} = m_0 n + {\cal O}(v^2)$, where
 $m_0$ is the particle rest mass, $n$ the rest frame number density
 and $v$ the rms particle velocity.  The off-diagonal terms of the
 energy-momentum tensor of the particle fluid average out to zero
 by summing over all particles with random velocities in all
 directions, whereas the $T^i_{\; i}\equiv -p$ components are
 proportional to the kinetic energy density, which for
 non-relativistic particles is negligible so that $p\ll \rho$.

 In the case of a `string gas' one can obtain an effective
 energy-momentum tensor in an analogous manner, by approximating
 the string network as a collection of straight string segments
 moving with average velocity $v$, and averaging over string
 orientations and directions of motion.  Let us first consider
 the relativistic case.  The effective energy-momentum tensor
 can be constructed by considering a straight string oriented
 in the $\hat {\bf z}$ direction say, and Lorentz-boosting its
 energy-momentum tensor in the $\pm \hat {\bf x}$ and $\pm \hat
 {\bf y}$ directions, then averaging and repeating the same
 procedure for strings oriented in the $\hat {\bf x}$ and $\hat
 {\bf y}$ directions \cite{Kolb_Turn}.  The result is:
 \be\label{T_fin}
  \langle T^\mu_{\; \nu} \rangle = \frac{\mu}{3L^2}
  \left(
   \begin{array}{cccc}
    3 \gamma^2 & 0 & 0 & 0 \\
    0 & (1-v^2\gamma^2) & 0 & 0 \\
    0 & 0 & (1-v^2\gamma^2) & 0 \\
    0 & 0 & 0 & (1-v^2\gamma^2) \\
   \end{array}
  \right) \, ,
 \ee
 where $\mu$ is the string tension, $L$ the average separation between
 nearby strings in the network and $\gamma=(1-v^2)^{-1/2}$ a Lorentz
 factor corresponding to $v$.  From (\ref{T_fin}) the equation of
 state can be read:
 \be\label{eos}
  -p \equiv \langle T^i_{\; i} \rangle = \frac{1}{3} ( \gamma^{-2}
  - v^2 ) \langle T^0_{\; 0} \rangle = \frac{1}{3} (1-2v^2) \langle
  T^0_{\; 0} \rangle \Rightarrow p=-\frac{1}{3} (1-2v^2) \rho \, .
 \ee

 A similar procedure can be followed for non-relativistic strings,
 which are generally expected to have much smaller string velocities.
 Indeed, for relativistic strings the constraint $\dot x^2+x^{\prime
 2}\equiv 0$ in the conformal gauge imposes that critical points on
 the string move with the speed of light, but for non-relativistic
 strings the physical string velocities can take any value.  One can
 thus obtain the equation of state for such a non-relativistic string
 gas by using Lorentz boosts with $\gamma=1$ or, alternatively, by
 performing transverse Galilean boosts instead.  The result is again
 $p=-\frac{1}{3}(1-2v^2) \rho$, but with the difference that one can
 safely assume $v\ll 1$, unlike the relativistic network case, where
 the strings oscillate relativistically at small scales, while there
 is no known mechanism which is efficient enough to damp these
 excitations.  Indeed, Hubble damping is inefficient at scales much smaller
 than the horizon, and for large scales, of order the string correlation
 length, numerical simulations (see for example \cite{All_Shell})
 demonstrate that string segments move more slowly and coherently, but at
 speeds large enough to produce significant deviations from the
 equation of state $w\equiv p/\rho=-1/3$.

 Note that one can apply an analogous procedure for strings which
 have a tension $T$ smaller than their mass per unit length $\mu$
 ($T<\mu$).  In this case the resulting equation of state is:
 \be\label{eos_Tmu}
  p=-\frac{1}{3} [ T/\mu(1-v^2) - v^2 ] \rho = -\frac{1}{3} [
  v_0^2 - (1+v_0^2) v^2 ] \rho \, ,
 \ee
 where we have defined the `sound speed' along the string $v_0 =
 \sqrt{T/\mu}$.  Equation (\ref{eos_Tmu}) can in general lead to both
 positive or negative equation of state with $p>-\rho/3$.  This is in
 contrast to vacuum (non-interacting) cosmic strings with $\mu=T$, where
 the rms speed does not exceed $1/\sqrt{2}$ so the equation of state is
 nonpositive (\ref{eos}) with  $p \ge -\rho/3$.  However, this is to be
 expected because in the limit $T\rightarrow 0$ the `string' describes a
 line-like structure of dust particles with $0<p\ll \rho$.  In fact,
 taking $v_0\rightarrow 0$ in equation (\ref{eos_Tmu}) gives $p=\rho
 v^2/3$, or, in terms of the kinetic energy density $\rho_k$,
 \be\label{kin}
  p=\frac{2}{3} \rho_k \, ,
 \ee
 which is precisely the equation of state for a gas of
 non-relativistic particles, following from ordinary kinetic theory
 considerations.  In connection to the discussion of the previous
 sections, obtaining this kind of non-relativistic string from the
 Nambu-Goto action involves a rescaling of the longitudinal directions
 by different factors, the ratio of which determines the propagation
 speed $v_0$.

 Finally, note that this discussion only applies to a `perfect' gas
 of non-interacting strings.  String intercommutations typically result
 in the removal of energy from the network in the form of closed
 string loops, significantly altering the above picture.  Thus, a
 \emph{frustrated} string network, with $w\simeq -1/3$, $\rho \propto
 a^{-2}$ eventually dominates over matter or radiation, but turning
 on string interactions will result to a different equation of state.
 For abelian string networks, where interactions are efficient,
 the resulting scaling law is $\rho \propto t^{-2}$, where $t$ is
 cosmic time, which scales like radiation in the radiation era and
 like matter in the matter era.  The cosmological evolution of
 non-relativistic string networks, including the possible effects of
 string intercommutation will be discussed in the next section.

 \section{\label{VOS} Velocity Dependent One-Scale (VOS) Models}\setcounter{equation}{0}

 In this section we discuss analytic models for the evolution of
 macroscopic variables describing the large-scale properties of
 a string network.  We will first review results for relativistic
 strings and then construct a macroscopic evolution model for
 non-relativistic strings, based on the action (\ref{S_NR}).
 To set up the physical picture we briefly summarise Kibble's
 one-scale model \cite{Kibble}, which captures the basic qualitative
 features of network evolution.

 Monte-Carlo simulations of cosmic string formation suggest that 
 to a good approximation the strings have the shapes of random walks 
 at the time of formation \cite{Vach_Vil}.  Such `Brownian' strings 
 can be described by a characteristic length $L$, which determines 
 both the typical radius of curvature of strings and the typical 
 distance between nearby string segments in the network.  On average 
 there is a string segment of length $L$ in each volume $L^3$ and 
 thus the density of the cosmic string network at formation is
 \be\label{rho}
  \rho=\frac{\mu L}{L^3}=\frac{\mu}{L^2} \,,
 \ee
 where $\mu$ is the string mass per unit length, which for
 relativistic strings is equal the `tension' $T_R$ appearing
 in the Lagrangian.  Assuming that the strings are simply stretched
 by the cosmological expansion we have $\rho \propto a(t)^{-2}$.
 This decays slower than both matter ($\propto a^{-3}$) and radiation
 ($\propto a^{-4}$) energy densities and so such non-interacting
 strings would soon dominate the universe.

 Now consider the effect of string interactions.  As the network
 evolves, the strings collide or curl back on themselves creating
 small loops, which oscillate and radiatively decay. Via these
 interactions enough energy is lost from the network to ensure that
 string domination does not actually take place.  Each string segment
 travels on average a distance $L$ before encountering another nearby
 segment in a volume $L^3$. Assuming relativistic motion ($v\approx 1$)
 and that the produced loops have an average size $L$, the
 corresponding energy loss is given by $\dot\rho_{\rm{loops}}\approx
 L^{-4} \mu L$. The energy loss rate equation is therefore
 \be\label{rholoss}
  \dot\rho\approx -2\frac{\dot a}{a}\rho - \frac{\rho}{L}\,.
 \ee
 Equation (\ref{rholoss}) has an attractor `scaling' solution in
 which the characteristic length $L$ stays constant relative to
 the horizon $d_H\sim t$ \cite{Kibble}.  The approach of string
 networks to a scaling regime has been verified by high-resolution
 simulations \cite{BenBouch, All_Shell}.

 Equation (\ref{rholoss}) was derived on physical grounds and it only
 captures the basic processes involved in string evolution, namely
 the stretching and intercommuting of strings. It does not take into
 account other effects like the redshifting of string velocities due
 to Hubble expansion.  In fact, it neglects completely the evolution
 of string velocities, making the crude approximation that they
 remain constant throughout cosmic history.  However, we can construct
 a more accurate Velocity-dependent One-Scale (VOS) model, based on
 the Nambu-Goto action (\ref{nambu}).

  \subsection{\label{rel_VOS}Relativistic Strings}

  The relativistic VOS model \cite{vos,vosk} extends Kibble's
  one-scale model, abandoning the constant string velocity
  approximation and introducing an extra variable, the rms velocity
  of string segments, whose dynamics is governed-as we will see-by
  a macroscopic version of the relativistic equations of motion
  (\ref{eom_x}).  Although the simple one-scale model captures most
  of the qualitative features of macroscopic string evolution, this
  correction is crucial for quantitative modelling.  Indeed, the
  average string velocity enters linearly in the loop production
  term, which provides the main energy loss mechanism of the string
  network, and so the evolution of string velocities significantly
  affects the string energy density.  The resulting VOS model is
  still very simple depending on only one free parameter\footnote{
  Strictly speaking there are two parameters in the VOS model, the
  loop production efficiency $\tilde c$ and the momentum parameter
  $k$.  For the second parameter, however, there exists a
  physically motivated ansatz (\ref{kans_R}), which expresses it
  in terms of the rms velocity $v(t)$.  Once this choice is made,
  one is only left with the freedom of tuning $\tilde c$ when
  trying to fit numerical simulations.} but, remarkably, it has
  been shown to accurately fit numerical simulation data throughout
  cosmic history \cite{vostests}.  We briefly sketch how the model
  is constructed from the microscopic equations of section
  \ref{rel_string}.  This will be useful for comparison to the 
  non-relativistic case. 

  Consider the relativistic string energy defined in section
  \ref{rel_string} (equation (\ref{string_energy})):
  \[
   E(\eta) = a(\eta) T_R \! \int \! \epsilon \, d\sigma
  \]
  and take the first derivative with respect to conformal time $\eta$.
  Using the equation of motion (\ref{eom_eps}) for $\epsilon$, one
  finds
  \be\label{E_dot_rel}
   \dot E = \frac{\dot a}{a} \left( 1 - 2 v^2 \right) E \, ,
  \ee
  where $v^2\!=\! \int \! \epsilon \dot{\bf x}^2 \, d\sigma /
  \! \int \! \epsilon \, d\sigma\! \equiv\! \langle \dot {\bf x}^2
  \rangle$ is the worldsheet average of the square of transverse
  velocities.  For a network of strings the energy density $\rho$
  is related to the total string energy by $E\propto \rho a(\eta)^3$.
  Therefore:
  \be\label{rho_dot}
   \frac{\dot \rho}{\rho} = \frac{\dot E}{E} - 3\frac{\dot a}{a}
   = -2 \frac{\dot a}{a} \left(1+v^2\right) \, .
  \ee
  To this we add a phenomenological term \cite{Kibble, book} describing
  the production of loops when strings collide and curl back on
  themselves.  The resulting network density evolution equation is:
  \be\label{rho_dot_full}
   \dot\rho = -2 \frac{\dot a}{a} \left(1+v^2\right) \rho - \tilde c
   \frac{v \rho}{L} \, ,
  \ee
  where $\tilde c$ is the loop production efficiency related to the
  integral of an appropriate loop production function over all
  relevant loop sizes \cite{book}.  This is treated as a free
  parameter which can be determined by comparison to numerical
  simulations.

  In the VOS model, the rms velocity $v$ appearing in equation
  (\ref{rho_dot_full}) is promoted to a dynamical variable whose
  evolution is given by a macroscopic version of the Nambu-Goto
  equation of motion (\ref{eom_x}).  This equation can be obtained
  by differentiating $v^2$ and eliminating $\ddot {\bf x}$ using
  the equation of motion.  This introduces the second spatial
  derivative ${\bf x}^{\prime\prime}$ which corresponds to string
  curvature and can be expressed in terms of the mean curvature
  radius of the network.  Differentiating $v^2$ and using equation
  (\ref{eom_eps}) we find:
  \be\label{v_square_dot}
   2v \dot v = \langle \dot {\bf x}^2 \rangle \dot{} = 2\langle
   \dot{\bf x}\cdot \ddot{\bf x} \rangle - 2\frac{\dot a}{a}
   \left( \langle \dot{\bf x}^2 \rangle^2 -\langle \dot{\bf x}^4
   \rangle \right) \, .
  \ee
  The second term is of purely statistical nature and has the effect
  of `renormalising' the coefficient of the $\frac{\dot a}{a} v^4$
  term which will find later.  It has been demonstrated numerically
  \cite{vos} to have small magnitude and thus can be neglected.

  Keeping only the first term and using the equation of motion for
  ${\bf x}$ we find:
  \be\label{v_vdot}
   v\dot v = \frac{\int\! \dot{\bf x} \cdot {\bf x}^{\prime\prime}
   \epsilon^{-1} \, d\sigma}{\int\!\epsilon\, d\sigma} + \frac{\int\!
   \dot{\bf x} \cdot {\bf x}^\prime (\epsilon^{-1})^\prime \,
   d\sigma}{\int\!\epsilon \, d\sigma} - 2 \frac{\dot a}{a} \left(
   \langle \dot{\bf x}^2 \rangle - \langle \dot{\bf x}^4 \rangle
   \right) \, .
  \ee
  The second term vanishes due to the gauge condition $\dot{\bf x}
  \cdot{\bf x}^\prime = 0$.  Further, within our approximations
  $\langle \dot{\bf x}^4 \rangle\simeq \langle \dot{\bf x}^2
  \rangle^2$ so the third term becomes $2\frac{\dot a}{a} v^2
  (1-v^2)$.  For the first term we need to express ${\bf
  x}^{\prime\prime}$ in terms of the local curvature vector.
  We define
  \be\label{ds}
   ds = \sqrt{ {\bf x}^{\prime2} } d\sigma = \sqrt{1-\dot{\bf x}^2}
   \epsilon d\sigma
  \ee
  and the physical (local) radius of curvature by
  \be\label{R}
   \frac{d^2 {\bf x}}{ds^2}=\frac{a(\eta)}{{\cal R}} \hat{\bf u} \, ,
  \ee
  where $\hat{\bf u}$ is a unit vector.  Then:
  \be\label{x_pp}
   {\bf x}^{\prime\prime} = \frac{d^2{\bf x}}{d\sigma^2}
   = {\bf x}^{\prime2} \frac{d^2 {\bf x}}{ds^2} + {\bf x}^\prime
   \frac{d \sqrt{{\bf x}^{\prime2}}}{ds}
  \ee
  Due to the constraint $\dot{\bf x} \cdot{\bf x}^\prime = 0$ the
  second term vanishes on dotting with $\dot{\bf x}$ so we have:
  \be\label{xdot_x_pp}
   \int\! \dot{\bf x} \cdot {\bf x}^{\prime\prime} \epsilon^{-1} \,
   d\sigma =
   \int\! \dot{\bf x}
   \cdot \frac{d^2 {\bf x}}{ds^2} ( 1 - \dot{\bf x}^2 ) \epsilon \,
   d\sigma = a(\eta) \langle (\dot{\bf x} \cdot \hat{\bf u} ) ( 1 -
   \dot{\bf x}^2 ) / {\cal R} \rangle \int\! \epsilon \, d\sigma \, .
  \ee
  We define the momentum parameter $k$ \cite{vosk} by the equation:
  \be\label{k}
   \langle (\dot{\bf x} \cdot \hat{\bf u} ) ( 1 - \dot{\bf x}^2 )
   / {\cal R} \rangle = \frac{kv}{{\cal R}} (1-v^2) \, ,
  \ee
  where ${\cal R}$ is now the average string radius of curvature,
  numerically close to the correlation length $L$ for Brownian networks
  \cite{book, vos, Aust_Cop_Kib}.  With this definition, equation
  (\ref{v_vdot}) becomes:
  \be\label{dv_dtau}
   \dot v = \frac{a(\eta)}{{\cal R}}k(1-v^2) - 2\frac{\dot a}{a}v(1-v^2) \, .
  \ee
  Changing to cosmic time $t$, with $dt = a d\eta$ and
  $\dot{}=a\frac{d}{dt}$ we finally obtain:
  \be\label{dv_dt}
   \frac{dv}{dt} = (1-v^2) \left( \frac{k}{{\cal R}} - 2Hv \right) \, ,
  \ee
  where $H=a^{-1} \frac{da}{dt}$ is the Hubble parameter.  Note that,
  since
  \[
   v^2=\langle \dot{\bf x}^2 \rangle=\left\langle \left(
   \frac{d{\bf x}} {d\eta} \right)^2 \right\rangle = \left\langle
   \left( a \frac{d{\bf x}}{dt} \right)^2 \right\rangle \,
  \]
  and the physical coordinates ${\bf x}_{\rm phys}$ are given in
  terms of the comoving ones ${\bf x}$ by ${\bf x}_{\rm phys} = a
  {\bf x}$, the rms velocity $v$ has the interpretation of physical
  \emph{peculiar} velocity of string segments.  Equation (\ref{dv_dt})
  has therefore a clear physical meaning: the rms peculiar velocities
  of string segments are produced by string curvature and damped by
  cosmological expansion.

  The momentum parameter $k$ is a measure of the angle between the
  curvature vector and the velocity of string segments and thus it is
  related to the smoothness of the strings.  As $v$ increases towards
  relativistic values the accumulation of small-scale structure renders
  the strings wiggly.  Velocities become uncorrelated to curvature and
  $k$ decreases.  In particular it can be shown analytically that for
  flat space, where $v^2=1/2$, the momentum parameter vanishes for a wide
  range of known solutions \cite{vos, Carl_thesis}.

  An accurate ansatz for the momentum parameter $k$ for relativistic
  strings has been proposed in \cite{vosk}
  \be\label{kans_R}
   k = k(v) = \frac{2\sqrt{2}}{\pi}\frac{1-8 v^6}{1+8 v^6} \, ,
  \ee
  satisfying $k(1/\sqrt{2})=0$.

  Note that the fact that $v=1/\sqrt{2}$ in flat spacetime, can be
  shown analytically for closed loops only, but for long strings it
  is observed in numerical simulations \cite{book}. For expanding
  or contracting spacetimes, $v$ is less or greater than $1/\sqrt{2}$
  respectively.  Hence for an expanding universe, string velocities
  are subject to the constraint:
  \be\label{vconstr}
   v^2 \le \frac{1}{2} \, .
  \ee
  In a matter or radiation dominated universe, Hubble expansion
  is too weak to significantly reduce string velocities, which
  remain close to $1/2$ at short scales \cite{book}.  This
  limitation does not apply to non-relativistic strings.

  \subsection{\label{NRVOS}Non-Relativistic Strings}

  For the non-relativistic string the energy of the excitations is
  given by (see Appendix \ref{NR_EM_tensor}):
  \be\label{E_exc}
   E_{\rm exc}=a(\eta) \frac{1}{2} \! \int \!  d\sigma \left(
   \mu \dot{\bf X}^2 + \mu \lambda^{-2} {\bf X}^{\prime 2}
   \right)=a(\eta)^{-1}\,{\cal P}_0  \, ,
  \ee
  where ${\bf X}$ are the \emph{transverse} string coordinates and
  we have defined the tension $\mu\!=\!\lambda T_0$.  To that we must
  add the string mass
  \be\label{E_0}
   E_0 = a(\eta) \mu \! \int \!  d\sigma \, ,
  \ee
  so that the total energy is:
  \be\label{E_tot}
   E=E_0 + E_{\rm exc}=a(\eta) \mu \! \int \!  d\sigma + a(\eta)^{-1}
   \,{\cal P}_0 \, .
  \ee
  Then, differentiating with respect to conformal time ( $\dot {}\
  \! = \frac{d}{d\eta}$), we have:
  \bq
   \dot E &=& \frac{\dot a}{a} E_0 + (a^{-1}{\cal P}_0)\dot{}
   = \frac{\dot a}{a} E_0 - \frac{\dot a}{a} E_{\rm exc} + a^{-1}
   \,\dot{\cal P}_0 \nonumber \\
   &=& \frac{\dot a}{a} \left(1 + \frac{1}{2} W^2 -
   \frac{3}{2} V^2 \right) E_0 \label{E0dot_E0} \, ,
  \eq
  where we have used equations (\ref{P_0_gauge}), (\ref{p_0_dot})
  and defined the rms quantities:
  \be\label{V}
   V^2= \frac{\int\! d\sigma \dot {\bf X}^2}{\int \! d\sigma} \equiv
   \langle \dot {\bf X}^2 \rangle  \, ,
  \ee
  and
  \be\label{W}
   W^2 = \frac{\int\! d\sigma \lambda^{-2} {\bf X}^{\prime 2}}{\int
   \! d\sigma} \equiv \langle \lambda^{-2} {\bf X}^{\prime 2}\rangle
   = \langle (\pd_{x^1}{\bf X})^2
   \rangle \, ,
  \ee
  corresponding to the average velocity of string segments and the
  average magnitude of string tangent vectors.  The latter quantity
  parametrises small-scale perturbations on the string, $W=0$ 
  corresponding to strings which are straight at scales smaller 
  than the correlation length\footnote{With this interpretation, 
  one expects that $W$ should have the effect of reducing the 
  effective radius of curvature of the network.  As we will see 
  later, this is indeed the case.}.  Thus, the term $W^2/2$ in 
  equation (\ref{E0dot_E0}) corresponds to the average elastic 
  energy due to short-scale string deformations.  In the 
  non-relativistic limit one has $W^2\ll 1$.    

  Defining the energy density $\rho \propto E a^{-3}$, and using
  \be\label{E_dot_E0}
   \frac{\dot E}{E_0} \simeq \frac{\dot E}{E} = \frac{\dot\rho}{\rho}
   + 3 \frac{\dot a}{a} \, ,
  \ee
  we find
  \be\label{rho_dot_NR}
   \dot\rho = -\frac{\dot a}{a} \left( 2 - \frac{1}{2} W^2
   + \frac{3}{2} V^2 \right) \rho - \tilde c V \frac{\rho}{L} \, ,
  \ee
  where we have included a phenomenological loop production term,
  as in the relativistic case.

  From (\ref{V}) we have
  \be\label{v_square_dot_NR}
   2 V \dot V = \langle \dot {\bf X}^2 \rangle^{\cdot} =
   2 \langle \dot{\bf X}\cdot \ddot{\bf X} \rangle - 2\frac{\dot
   a}{a}\left( \langle \dot{\bf X}^2 \rangle^2 -\langle \dot{\bf X}^4
   \rangle \right)
  \ee
  as before.  We neglect the statistical terms and using the
  non-relativistic equation of motion (\ref{eom_gauge}) we find:
  \be\label{vv_dot_NR}
   V\dot V = \frac{\int\!\dot{\bf X} \cdot \ddot{\bf X} \, d\sigma}
   {\int \! \, d\sigma} = \frac{\int\! \lambda^{-2} \dot{\bf X}
   \cdot {\bf X}^{\prime\prime}\, d\sigma}{\int \! \, d\sigma}
   - 2\frac{\dot a}{a} V^2
  \ee
  In order to express ${\bf X}^{\prime\prime}$ in terms of the
  string curvature vector we define:
  \be\label{ds_NR}
   ds = \sqrt{ 1 + (\pd_{x^1} {\bf X})^2 } dx^1=
   \lambda \sqrt{ 1 + \lambda^{-2} {\bf X}^{\prime2} } d\sigma
  \ee
  and the physical radius of curvature:
  \be\label{R_NS}
   \frac{d^2 {\bf Y}}{ds^2}=\frac{a(\eta)}{{\cal R}}\hat{\bf u}\,,
  \ee
  where we have introduced the 3-vector ${\bf Y}\!=\!\left(x^1,{\bf
  X}\right)$ and a unit 3-vector $\hat{\bf u}$.  Now:
  \be\label{X_pp_NR}
   {\bf X}^{\prime\prime} = \frac{d^2{\bf X}}{d\sigma^2} =
   \lambda^2 \left(1+\lambda^{-2}{\bf X}^{\prime2}\right) \frac{d^2
   {\bf X}}{ds^2} + \lambda {\bf X}^\prime \, \frac{d \sqrt{1+
   \lambda^{-2}{\bf X}^{\prime2}}}{ds}
  \ee
  In this case, the second term will not cancel on dotting with $\dot
  {\bf X}$, because $\dot {\bf X} \cdot {\bf X}^\prime \ne 0$ for the
  non-relativistic string.  Instead we have two terms:
  \be\label{Xdot_X_pp_NR}
   \lambda^{-2} \int\! \dot{\bf X} \cdot {\bf X}^{\prime\prime} \,
   d\sigma = \int\! \dot{\bf X} \cdot \frac{d^2 {\bf X}}{ds^2}
   \left(1 + \lambda^{-2}{\bf X}^{\prime2}\right)  \, d\sigma +
   \lambda^{-2} \int\! \dot{\bf X} \cdot{\bf X}^\prime  \left(\ln
   \sqrt{1 + \lambda^{-2}{\bf X}^{\prime2}} \right)^\prime
   d\sigma \, .
  \ee
  For the first term we note that, since $\dot{\bf X}$ is normal
  to $(x^1,{\bf 0})$ in Cartesian coordinates,
  \be\label{Xdot_curv}
   \dot{\bf X} \cdot \frac{d^2 {\bf X}}{ds^2} = \dot{\bf X} \cdot
   \frac{d^2{\bf Y}}{ds^2}
  \ee
  and so we can use equation (\ref{R_NS}) to write:
  \be\label{int_Xdot_curv}
   \int\! \dot{\bf X} \cdot \frac{d^2 {\bf X}}{ds^2} \left(1
   + \lambda^{-2}{\bf X}^{\prime2}\right)  \, d\sigma =
   a(\eta) \frac{k V}{{\cal R}} (1 + W^2) \!\int\! d\sigma\, .
  \ee
  Here, in analogy to the relativistic case, we have defined a
  momentum parameter $k$ by:
  \be\label{kdef_NR}
   \left\langle \left(1 + \lambda^{-2}{\bf X}^{\prime2}\right)
   (\dot{\bf X} \cdot \hat{\bf u}) /{\cal R} \right\rangle =
   \frac{k V}{{\cal R}} (1 + W^2) \, .
  \ee

  For the second term in (\ref{Xdot_X_pp_NR}) we have:
  \bq
   && \lambda^{-2} \int\! \dot{\bf X} \cdot{\bf X}^\prime  \left(\ln\sqrt{1
   + \lambda^{-2}{\bf X}^{\prime2}} \right)^\prime d\sigma
   = \lambda^{-2} \int\! \dot{\bf X} \cdot{\bf X}^\prime \frac{{\bf X}^\prime
   \cdot {\bf X}^{\prime\prime}\lambda^{-2}}{1+ \lambda^{-2}{\bf
   X}^{\prime2}} \, d\sigma
   \nonumber \\
   && \ \ = \lambda^{-2} \int\! \left(\dot{\bf X} \cdot{\bf X}^\prime \right)
   \left({\bf X}^\prime \cdot \hat {\bf u}\right) \frac{a(\eta)}{{\cal R}}
   \, d\sigma + \lambda^{-3} \int\! \left(\dot{\bf X} \cdot{\bf X}^\prime
   \right){\bf X}^{\prime2} \frac{{\bf X}^\prime \cdot {\bf X}^{\prime
   \prime}\lambda^{-2}}{\left(1+ \lambda^{-2}{\bf X}^{\prime2}\right)^2} \,
   d\sigma \nonumber \\
   && \ \ = a(\eta) \frac{k^\prime V W^2}{{\cal R}} \!\int\! d\sigma
   + {\cal O}(VW^4) \, ,
   \label{Xdot_Xp_NR}
  \eq
  where we have used equation (\ref{X_pp_NR}) and defined the
  parameter $k^\prime$ by:
  \be\label{k_prime}
   \left\langle \lambda^{-2} \left(\dot{\bf X} \cdot{\bf X}^\prime
   \right) \left({\bf X}^\prime \cdot \hat {\bf u}\right)/{\cal R}
   \right \rangle = \frac{k^\prime V W^2}{\cal R}
  \ee
  Putting all the terms together, equation (\ref{vv_dot_NR}) can be
  rewritten (in terms of cosmic time $t$) as:
  \be\label{dV_dt}
   \frac{dV}{dt}=\frac{1}{{\cal R}} \left( k + k^{\prime\prime} W^2
   \right) - 2HV \, ,
  \ee
  with
  \be\label{kpp}
   k^{\prime\prime}\equiv k + k^{\prime}\, .
  \ee

  Equations (\ref{rho_dot_NR}), (\ref{dV_dt}) form the
  Non-Relativistic Velocity dependent One-Scale (NRVOS) model.
  In principle one should consider $W$ as a third dynamical variable
  and try to derive an evolution equation, as in the case of $V$.
  As a first approximation we will assume that time variations in $W$
  do not have a significant impact, $W$ remaining always small, and
  we will treat it as a constant parameter.  This approximation will
  be tested in the next section, where we will solve the NRVOS
  equations numerically, for different choices of the $W$ parameter.

  Finally, one comment is in order regarding the magnitude of the 
  parameter $k^\prime$.  From its definition in equation (\ref{Xdot_Xp_NR})
  one expects $k^\prime\ll k$.  Indeed, $k^\prime$  measures the
  average value of $(\dot{\bf X}\cdot{\bf X}^\prime)({\bf X}^\prime \cdot
  \hat{\bf u})$ the first factor of which contains uncorrelated vectors,
  while for the second factor string tangents will generally be normal to
  the local curvature vector.  On the other hand $k$ corresponds to the
  average value of $\dot{\bf X} \cdot \hat {\bf u}$ and these two vectors
  are correlated, at least for smooth strings/small excitation velocities.
  Given that the $k^{\prime\prime}$ term in (\ref{dV_dt}) is already
  suppressed by a factor ${\cal O}(W^2)$ it is a good approximation to 
  set $k^{\prime\prime}\simeq k$.  Then, $W$ has the effect of  
  `renormalising' the effective radius of curvature ${\cal R} 
  \rightarrow {\cal R}/(1+W^2)$ (or equivalently the momentum
  parameter $k\rightarrow k(1+W^2)$), as may be expected from its
  interpretation as a short-scale structure parameter.

 \section{\label{RelVnonRel}Relativistic vs Non-Relativistic Network Evolution}\setcounter{equation}{0}

 In this section we solve numerically the NRVOS equations for a
 non-relativistic string network and compare to the relativistic
 case.  The naive expectation is that non-relativistic networks
 are denser than their relativistic counterparts because the small
 string velocities reduce the effect of the loop production term.
 Physically, the transverse excitations on strings are non-relativistic
 so fewer loops are produced per unit time due to string
 self-intersections.  Long string segment interactions are also
 suppressed due to the low collision rate corresponding to
 small velocities.

 To close the NRVOS equations we need to specify an ansatz for the
 non-relativistic momentum parameter $k$.  For a velocity dependent 
 model like the one we have developed, it is not consistent to treat 
 $k$ as a constant parameter.  Further, in the relativistic case, its 
 dependence on the rms velocity $v$ (equation (\ref{kans_R})) is   
 important in determining the scaling values of the network variables.  
 The functional dependence of the momentum parameter on $v$ can be 
 obtained by considering `curvature' and `bulk' contributions to string  
 velocities, as explained in Ref.~\cite{vosk}.  Following the discussion 
 in that reference we take:    
 \be\label{k_NR}
  k(v) = k_0 (1-v^2) \, ,
 \ee
 where $k_0$ is a constant.  This has the same functional dependence 
 as the low-velocity limit of $k(v)$ in Ref.~\cite{vosk}, but here 
 we have left the overall normalisation $k_0$ as a free parameter.  
 This reflects the fact that the non-relativistic string limit is 
 not merely a low-velocity one.  There is a difference between
 slowly moving, straight, relativistic strings and wiggly, 
 non-relativistic strings.  The defining property of the non-relativistic
 string is that its transverse excitations be Galilei, as opposed to 
 Lorentz, invariant.  The difference between relativistic and 
 non-relativistic strings is in the transverse perturbations. 
 In an effective description, non-relativistic strings can be 
 thought of as having a short wavelength cut-off on the string 
 excitations.  As a result, arbitrarily small-wavelength relativistic 
 perturbations are not excited and this translates into a reduced 
 curvature parameter normalisation $k_0$.  The string can be 
 thought of as a massive rigid rod, but with tension $T$ equal to its 
 mass per unit length $\mu$ \footnote{Relativistic invariance in the 
 longitudinal directions implies that the waves along the string travel 
 at the speed of light $c$ \cite{JGom_Kam_Town}.  Note the difference 
 to the other notion of non-relativistic string with $T<\mu$ and 
 longitudinal speed $v<c$.}.  In analogy to the 
 relativistic case, where the overall normalisation was determined 
 by comparison to a known analytic solution \cite{vosk}, $k_0$ can 
 be obtained in the non-relativistic case by comparison to a given 
 model of non-relativistic string.  In the general discussion below 
 we will simply treat it as a free parameter and examine its effect 
 on the network evolution.  

 Equations (\ref{rho_dot_NR}), (\ref{dV_dt}) and (\ref{k_NR}) have
 been solved numerically for a range of parameters $k_0$ and
 $W$.  This was done by rewriting equation (\ref{rho_dot_NR})
 in terms of the correlation length $L=\sqrt{\mu/\rho}$ and then
 introducing a function $\gamma(t)=L/t$.  Under the assumption
 $L\simeq {\cal R}$, the resulting equation for $\gamma(t)$ together
 with (\ref{dV_dt}) form a non-autonomous system of coupled ODE's,
 which can be integrated numerically.  During matter or radiation 
 domination, this system has an attractor solution in which both
 $\gamma(t)$ and $v(t)$ tend to constant values (scaling).  Here, 
 we present numerical results for a radiation dominated universe.  

\begin{figure}
   \begin{center}
   \includegraphics[width=2.7in,keepaspectratio]{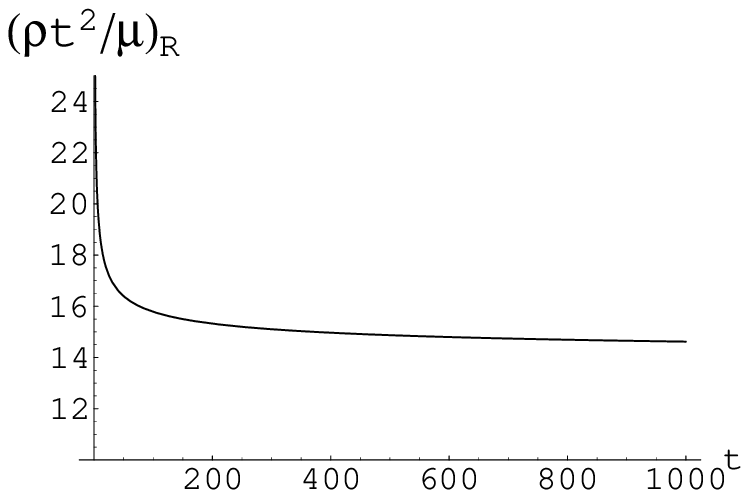}
   \includegraphics[width=2.7in,keepaspectratio]{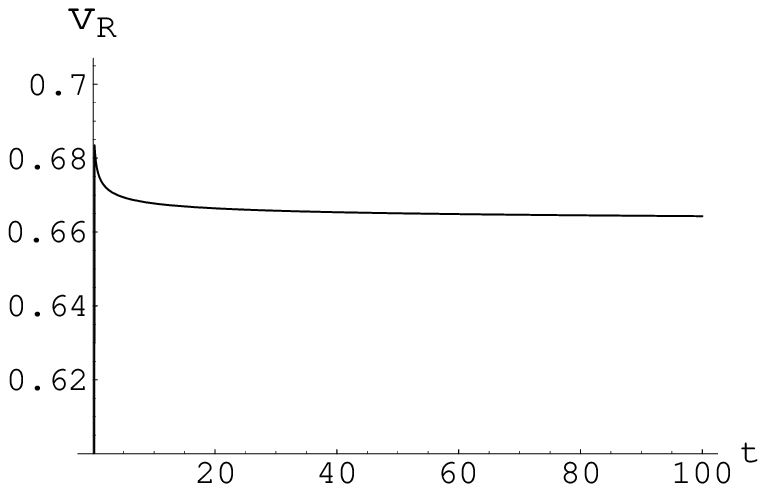}
   \includegraphics[width=2.7in,keepaspectratio]{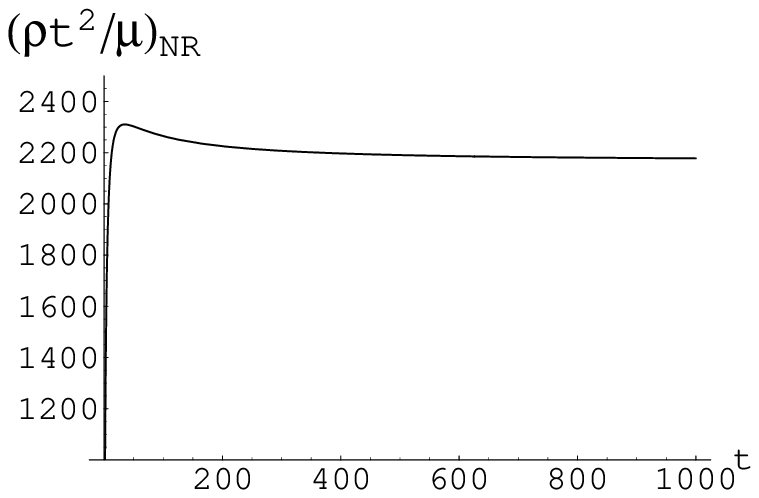}
   \includegraphics[width=2.7in,keepaspectratio]{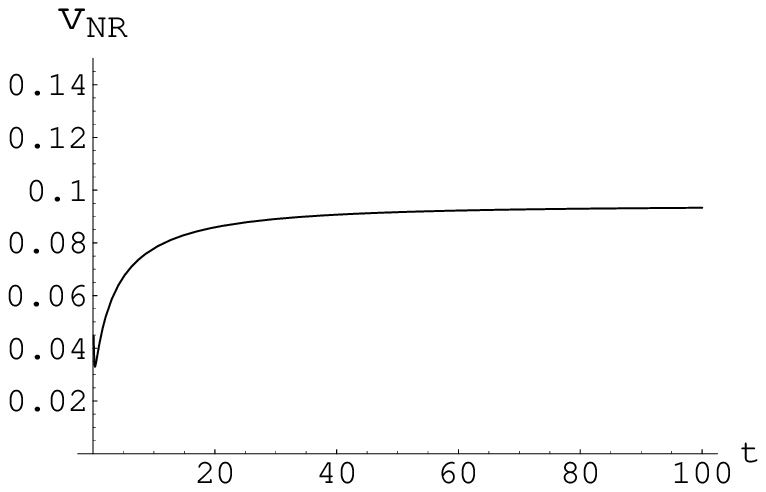}
 \end{center}
   \caption{\label{fig_comparison}  Relativistic versus
   non-relativistic network evolution.  Non-relativistic networks
   evolve to slower and much denser scaling configurations than
   their relativistic counterparts.  Here, we have plotted the
   evolution of the normalised string density and rms velocity
   for a relativistic network and a non-relativistic one with
   $V\simeq 0.1$.}
  \end{figure}

 In Fig.~\ref{fig_comparison} we plot the evolution of the string
 energy density and rms velocity for both a non-relativistic
 and a relativistic string network, that is, the solution of
 equations (\ref{rho_dot_NR}), (\ref{dV_dt}), (\ref{k_NR})
 in the former case and (\ref{rho_dot_full}), (\ref{dv_dt}),
 (\ref{kans_R}) in the latter.  To highlight the effect of
 non-relativistic velocities, we have chosen a value of the
 parameter $k_0$ which gives a scaling value of $V\simeq 0.1$ and taken
 $W<V$.  We have also assumed that both networks have the same loop
 production efficiency parameter $\tilde c$ and chosen the value
 $\tilde c=0.23$, suggested by relativistic network simulations
 \cite{vos}.  As expected, the non-relativistic network has a much
 higher scaling string density compared to the non-relativistic one.
 Of course, non-interacting strings ($\tilde c=0$) do not converge
 to a scaling solution.

\begin{figure}
  \begin{center}
   \includegraphics[width=2.7in,keepaspectratio]{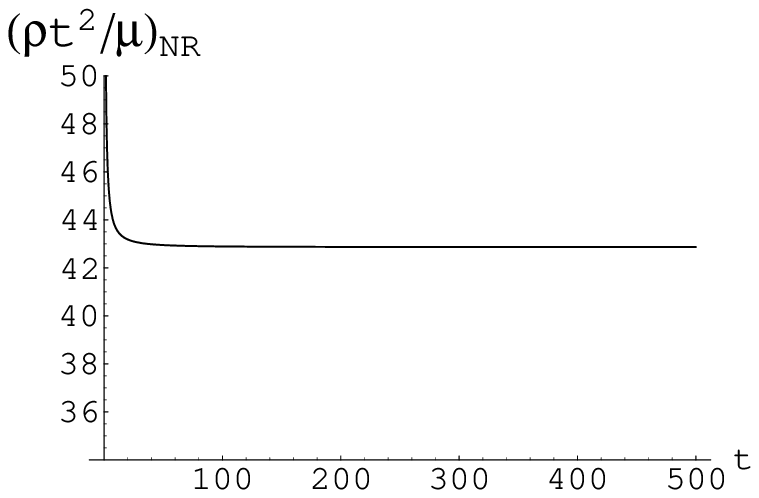}
   \includegraphics[width=2.7in,keepaspectratio]{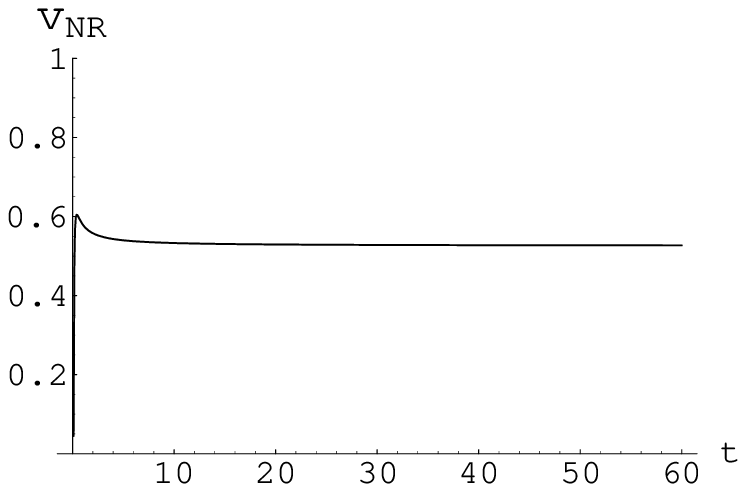}
   \includegraphics[width=2.7in,keepaspectratio]{rhoNR_v01.eps}
   \includegraphics[width=2.7in,keepaspectratio]{vNR_v01.eps}
   \includegraphics[width=2.7in,keepaspectratio]{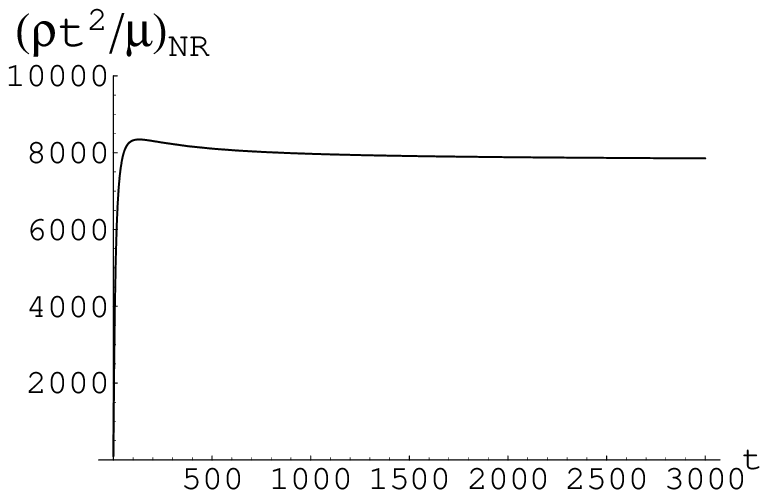}
   \includegraphics[width=2.7in,keepaspectratio]{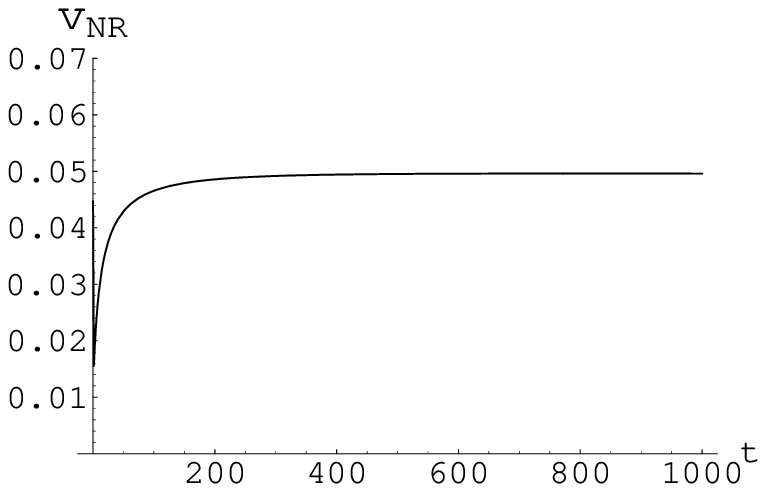}
  \end{center}
   \caption{\label{fig_NR_v0} Evolution of normalised string density
   and rms velocity for a non-relativistic network with constant $W<V$
   for different choices of the parameter $k_0$.  Reducing the value
   of $k_0$ results to lower rms string velocities, which in turns
   implies a slower rate of string interactions.  This results in a
   dramatic enhancement of string network density.}
  \end{figure}

 We now explore the dependence of non-relativistic string evolution
 on the parameters $k_0$ and $W$.  In Fig.~\ref{fig_NR_v0} we plot
 the normalised string density $\rho t^2/\mu=\gamma^{-2}$ and the
 rms string velocity $V$ as functions of cosmic time $t$ for different
 choices of $k_0$ producing string velocities $0<V<1$.  We have assumed
 a constant value of $W<V$, but below we will consider the effect of
 varying $W$ also, allowing for $W>V$.  It is apparent from
 Fig.~\ref{fig_NR_v0} that the rms string velocities are controlled
 by the parameter $k_0$.  Reducing $k_0$ leads to smaller $V$, which
 in turn implies a higher string density, due the reduced energy loss
 term.  The fact that the scaling value of the rms velocity is not 
 universal for non-relativistic strings, but instead depends on the
 parameter $k_0$, is not surprising.  In the relativistic case, there 
 is a distinct upper speed limit $c=1$ and the relativistic constraint 
 implies that the rms velocities are smaller than, but not far off,  
 $1/\sqrt{2}$ (see for example Ref.~\cite{book}).  On the other hand,
 in any non-relativistic theory velocities are unbounded.  

\begin{figure}
   \begin{center}
  \includegraphics[width=2.7in,keepaspectratio]{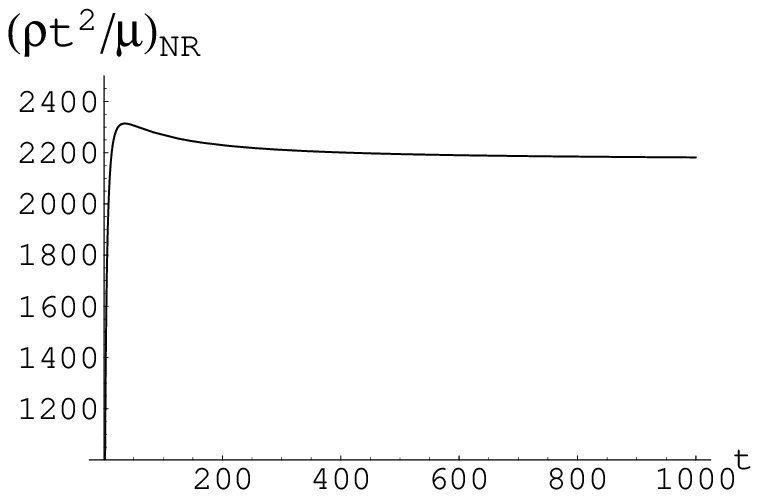}
  \includegraphics[width=2.7in,keepaspectratio]{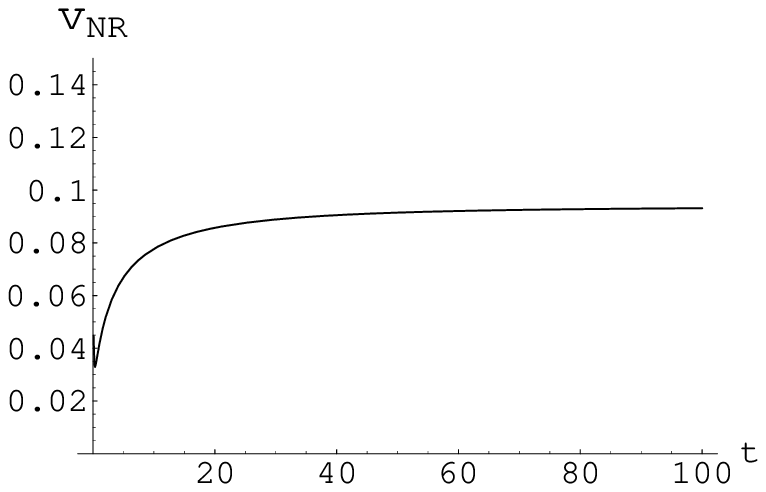}
  \includegraphics[width=2.7in,keepaspectratio]{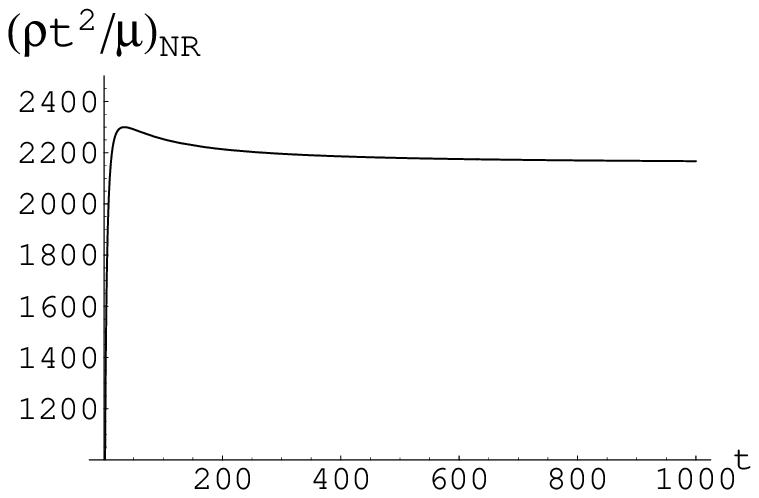}
  \includegraphics[width=2.7in,keepaspectratio]{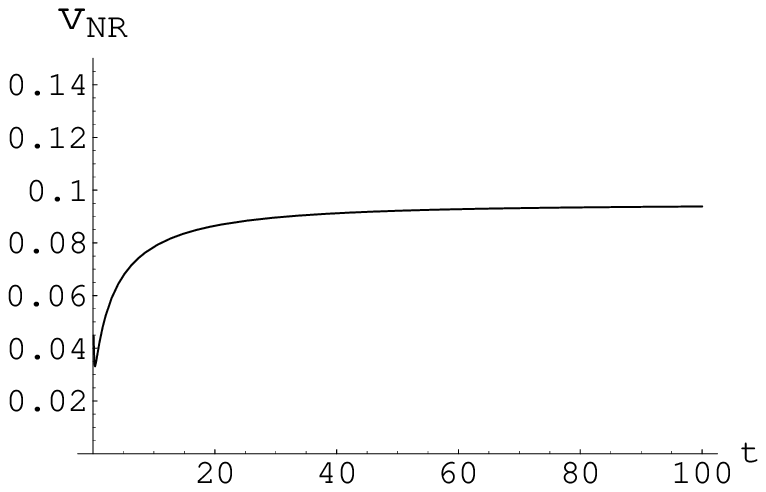}
  \includegraphics[width=2.7in,keepaspectratio]{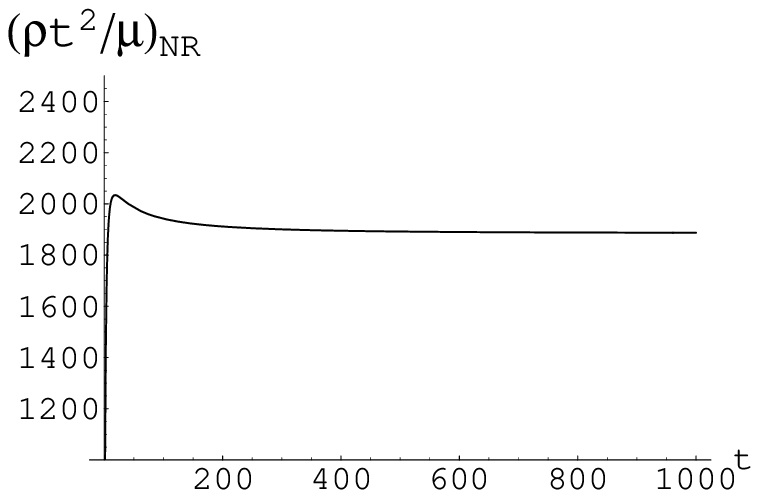}
  \includegraphics[width=2.7in,keepaspectratio]{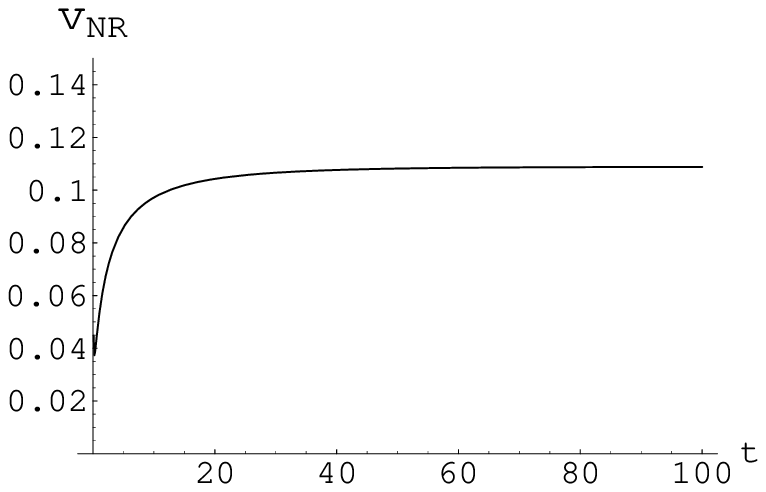}
   \end{center}
   \caption{\label{fig_W} Dependence of normalised string density
   on the parameter $W$ for a network with $V\simeq 0.1$.
   The plots correspond to $W=0, 0.1$ and $0.5$ respectively.
   Increasing $W$ does not significantly alter the scaling density
   until $W$ becomes greater than $V$.  For $W=0.5=5V$, the scaling
   is reduced by $10\%$, so it remains two orders of magnitude greater
   than that of relativistic strings.}
  \end{figure}

 We then consider the impact of varying the parameter $W$.  Looking
 at the first term of equation (\ref{rho_dot_NR}), which describes
 dilution due to cosmic expansion, one observes that $W^2$ and $V^2$
 appear with opposite signs, so a large $W$ could counterbalance
 (or even reverse) the effect of $V$ on this term.  However, if
 both $V,W\ll 1$ they play no significant role in that term.  Thus,
 one only needs to check the case $W>V$ when $V,W$ are not negligible.
 Fig.~\ref{fig_W} shows the time evolution of $\rho$ for a choice of
 $k_0$ leading to $V\simeq 0.1$, for the cases $W=0, 0.1, 0.5$.
 The first two figures show identical evolutions, even though
 in the second one $W\simeq V$.  In the third case, however, where
 $W^2=25 V^2$ the effect of $W$ counterbalances that of $V$ in the
 dilution term of (\ref{rho_dot_NR}), resulting in an appreciable
 reduction of the string scaling density, at the $10\%$ level.
 Since the most important impact of string velocities is through
 the loop production term of (\ref{rho_dot_NR}), the basic prediction
 of the model, which is a dramatic enhancement of the string scaling
 density (Fig. \ref{fig_comparison}), remains robust.

 \section{\label{discuss}Discussion}\setcounter{equation}{0}

 So far we have studied the dynamics and macroscopic evolution of
 non-relativistic strings in some generality, without discussing
 any specific setup in which they could be relevant.  However,
 non-relativistic string-like objects arise in several contexts
 and have been considered before in the literature.

 For example, Ref.~\cite{Mart_Moor_Shell} studied non-relativistic
 vortex-strings with motivations from both cosmology \cite{book} and
 condensed matter physics \cite{Schwarz85, Schwarz88}.  In that
 reference, the non-relativistic limit was taken at the level of the
 equations of motion by requiring small string velocities $\dot X^2\ll
 1$.  Here, we have taken the non-relativistic limit at the level of
 the string action but this involved a rescaling procedure which
 corresponds to having both $\dot X^2\ll 1$ and $(\pd X/\pd \zeta)^2
 \ll 1$, where $\zeta$ is the physical length along the string.
 The non-relativistic evolution model we have developed in section
 \ref{NRVOS} can be applied to the condensed matter context considered
 in \cite{Mart_Moor_Shell} by introducing a friction term relevant
 to that situation.  Adding this term and setting $\dot a/a=0$
 equation (\ref{rho_dot_NR}), expressed in terms of the correlation
 length, reads:
 \be\label{}
 2\frac{dL}{dt}=\tilde c V + \frac{L}{\ell_d} V^2\, ,
 \ee
 as in Ref.~\cite{Mart_Moor_Shell}, where $\ell_d$ is the relevant
 damping length scale.  The velocity evolution equation is also modified 
 by the addition of a friction term $-\ell_d/L$, again as in 
 Ref.~\cite{Mart_Moor_Shell}.  The system has a solution with $L\propto 
 t^{1/2}$, which is actually observed experimentally for defects in 
 condensed matter systems and liquid crystals \cite{Mermin, SalVol, ChDTY}.

 In cosmology, slowly-evolving string networks have been invoked
 in order to obtain a negative equation of state \cite{SperPen}.
 Bucher \& Spergel \cite{BuchSper} have proposed a Solid Dark Matter
 (SDM) model, which could be realised in terms of a frustrated string
 or domain wall \cite{BatBuchSper} network.  Rigidity and stability in
 this scenario have been studied in Ref.~\cite{BatCarChMo}.  More recently,
 a string network of the SDM kind was revived \cite{Alexand} in an
 attempt to explain the flat rotational curves and the Tully-Fisher
 relation observed in galaxies, which were the main motivation for
 the development of MOND\footnote{For a recent review on the MOND
 scenario see Ref.~\cite{MOND}.} theories.  The fundamental difficulty
 \cite{BatCarChMo} with the SDM scenario is to explain how an essentially
 non-relativistic network can naturally arise from an initial tangle
 of (relativistic) Nambu-Goto strings like the ones believed to be
 produced in cosmological phase transitions.  Indeed, Hubble damping
 is inefficient at subhorizon scales \cite{book} and there is no known
 mechanism efficient enough to damp the relativistic short-scale excitations
 on strings.  These affect the equation of state through the velocity
 dependent term in equation (\ref{eos}), leading to $w>-1/3$\footnote{
 One could argue that the velocity which enters the equation of state
 is the coherent string velocity at the scale of the string
 correlation length rather than the rms short-scale velocity.  While
 it is true that the coherent velocities are typically smaller,
 numerical simulations \cite{All_Shell} suggest $v_{\rm coh}\simeq 0.15$
 so one still expects significant departure from $w=-1/3$.  Furthermore,
 small-scale structure has the effect of `renormalising' the string
 mass per unit length \cite{book, Mart_Shell_sss} and string tension
 so that equation (\ref{eos_Tmu}) should be used instead of (\ref{eos}).
 This also increases the value of $w$.}.  Further, numerical
 evidence is now accumulating supporting that scaling behaviour in
 field theory strings and domain walls is rather generic \cite{walls},
 so that frustrated networks seem hard to obtain.  On the other hand
 the analysis we did in section \ref{RelVnonRel} points towards a SDM
 picture for non-relativistic strings, where the above problems are
 not present.  Here, string velocities can be arbitrarily small and,
 as we saw in section \ref{RelVnonRel}, network densities are
 dramatically enhanced so that strings could even dominate the
 universe before scaling is reached.

 Note that the procedure for obtaining the non-relativistic string
 action (\ref{S_NR}) required at least one of the spatial directions
 to be compact.  If the action (\ref{S_NR}) is to be treated as a
 classical effective action this global property can be ignored, but
 if it is taken to describe a fundamental object, then the winding
 around a compact dimension is required at quantum level.  The fact
 that a consistent non-relativistic string theory based on the action
 (\ref{S_NR}) can be constructed \cite{jGom_Oog} allows one to take
 the view that there is a fundamental winding string obeying this
 action.  Then, a cosmological setup like that of sections \ref{NRVOS}
 and \ref{RelVnonRel} can still be considered as long as the
 compactification radius is larger than the horizon.  This possibility
 of having a universe with non-trivial topology is not observationally
 excluded.  Cosmological observations constrain the local geometry 
 as described by the metric to be nearly flat \cite{WMAP3}, but the
 global topology of spatial hypersurfaces need not be that of the
 covering space.  Indeed, topological identifications under
 freely-acting subgroups of the isometry group are allowed, and
 the WMAP sky maps appear to be compatible with finite flat topologies  
 with fundamental domain significantly greater than the distance to 
 the decoupling surface \cite{Luminet} (see also \cite{Cornish}).

 One can therefore imagine a situation where fundamental
 non-relativistic strings are wound around 1-cycles in a
 non-simply-connected universe, in a setup analogous to that
 of the Brandenberger-Vafa scenario \cite{BrandVaf}.  If the
 compactification radius is larger than the horizon, as required
 by cosmological observations, a network of such wound strings
 behaves like an open string network.  An analogous situation
 occurs in ordinary cosmic string simulations, where the
 network evolves in a periodic box and there is a class of long
 strings (determined mainly by initial conditions) which wind around
 the box.  As the universe expands these strings tend to straighten
 out and behave essentially non-relativistically \cite{Paul_private}.
 These strings are usually discarded as artifacts of the periodicity
 of the box, but in a universe of compact topology, such
 configurations can play a physical role.

 Finally, in theories with compact extra dimensions one has the
 possibility of non-relativistic strings winding 1-cycles in the
 internal space.  Analogous (but relativistic) objects have been
 considered in the context of brane inflation
 \cite{BarnBCS, MatsNecl, cycloops}, which are topologically trapped
 and behave like monopoles.  Although the copious production of such
 objects in the early universe is inconsistent with the existence of
 an early radiation era, there are regions in parameter space where
 they are allowed and in some cases can provide candidates for dark
 matter.  The situation of non-relativistic strings wrapping an internal
 dimension is qualitatively similar, but the corresponding energy
 spectrum is different than in the relativistic case.

 The outstanding question arising from the above is to what extent
 such non-relativistic strings are `natural' or `generic' objects
 in cosmology.  Even though non-relativistic strings exist in some
 part of the moduli space of string theory, there is at present no
 mechanism which produces them in a cosmological setup.  Nevertheless,
 it is clear that the non-relativistic string action and the VOS model
 developed here are applicable at least as effective descriptions of
 cosmic- and vortex-strings in certain situations.  Indeed, the action 
 we have considered is the only sensible non-relativistic limit, having 
 $T=\mu$, of the standard Nambu-Goto action, and is precisely the action 
 one obtains when considering the low energy dynamics of topological 
 defects in field theory.  The macroscopic NRVOS model based on this 
 action, provides a semi-analytic tool for the study of the cosmological
 evolution of non-relativistic strings.  Possible situations of 
 cosmological interest involving non-relativistic strings include strings 
 in de Sitter space, Solid Dark Matter, wound strings, etc, as discussed
 above.  Further, in a condensed matter application we have noted that 
 our model reproduces the correct scaling law, as experimentally observed.     

 It would be interesting to go one step further and perform numerical 
 simulations of string network evolution based on the non-relativistic 
 string action presented here.  The comparison of macroscopic string 
 evolution and small-scale structure to the relativistic case could 
 provide an independent means of probing the effect of small-scale 
 structure on string networks, which is an area of current interest 
 and active research.

\begin{acknowledgments}
We are grateful to Carlos Martins for reading the manuscript
and making valuable comments.   It is also a pleasure to thank 
Roberto Emparan, Jaume Garriga, Gary Gibbons, Jaume Gomis, Paul 
Shellard and Paul Townsend for discussions.  This work has been 
supported in part by the European EC-RTN project MRTN-CT-2004-005104, 
European Network on Random Geometry (ENRAGE) project 
MRTN-CT-2004-005616, MCYT FPA 2004-04582-C02-01 and 
CIRIT GC 2005SGR-00564. We would like to thank the Galileo Galilei 
Institute for Theoretical Physics for its hospitality and INFN 
for partial support during the completion of this work.

\end{acknowledgments}

\appendix

\section{\label{semicl} Non-Relativistic String from Semiclassical Approximation}

 In this section we derive the action (\ref{S_NR}) as a semiclassical
 expansion around the vacuum solution.  Non-Relativistic D-brane
 actions on AdS$_5\times S^5$ have been recently constructed
 \cite{semiclass} with this method, by considering the
 Dirac-Born-Infeld (DBI) action and expanding around a classical
 solution.  Here, we apply the method to the case of the Nambu-Goto
 action in an expanding FLRW background.

 We start from the Nambu-Goto action
 \be\label{S_nambu}
  S_{NG}=-T \! \int \! \sqrt{-\gamma}\, d^2\sigma
 \ee
 and write the induced metric $\gamma_{ij}=G_{MN}\pd_i x^M \pd_j
 x^N$ as
 \be\label{FLRW_viel}
  \gamma_{ij}=e^M_i e^N_j \eta_{MN} \, ,
 \ee
 where $e^M_i=a(x^0)\pd_i X^M$ is the (worldsheet induced) vielbein
 for the FLRW metric (\ref{FLRW_metric}).  We now consider the vacuum
 field configuration:
 \be\label{clacc_sol}
  x_0^M = \left\{
   \begin{array}{cll}
           \tau       &,&      M=0           \\
       \lambda\sigma  &,&      M=1           \\
             0        &,&  M=a\in (2,\ldots,D)
   \end{array}
  \right.
 \ee
 which is a solution of the equations of motion (\ref{eom}).  The only
 non-trivial vielbeins evaluated on the solution are the longitudinal
 ones
 \be\label{viel_long}
  e^{\; \mu}_{0\, i} = a(x^0)\pd_i X^\mu\, , \quad \mu=0,1 \, ,
 \ee
 the transverse ones $e^{\; a}_{0\, i}$ being zero.  The induced metric on
 the static solution is then
 \be\label{gamma_0}
  \gamma_{0\,ij}=e^{\, M}_{0\,i} e^{\, N}_{0\, j} \eta_{MN}
  = e^{\; \mu}_{0\,i} e^{\; \nu}_{0\, j} \eta_{\mu\nu}  \, .
 \ee
 Next we introduce transverse fluctuations around this solution,
 namely
 \be\label{fluct}
  x^a=X^a(\sigma^i) \, .
 \ee
 The induced metric can then be written as the sum of the static
 solution plus a piece quadratic in the fluctuations:
 \be\label{g_0_g_2}
  \gamma_{ij} = \gamma_{0\, ij} + \gamma_{2\, ij} \, ,
 \ee
 where
 \bq
  &&\gamma_{0\,ij} = a(x^0)^2 \,{\rm diag}\left(-(\pd_\tau x^0)^2, 
     \,(\pd_\sigma x^1)^2 \right) = a(x^0)^2 \,{\rm diag}(-1,\lambda^2)   
     \label{g_0} \\
  &&\gamma_{2\,ij} = a(x^0)^2 \pd_i X^a \pd_j X^b \delta_{ab}
     \label{g_2} \, .
 \eq
 Expanding $\sqrt{-\gamma}$ we have:
 \bq
  \sqrt{-{\rm det}\gamma}&=&\sqrt{-{\rm det}(\gamma_0+\gamma_2)}
  =\sqrt{-{\rm det}[\gamma_0(1+\gamma_0^{-1}\gamma_2)]}\nonumber \\
  &=& \sqrt{-{\rm det}\gamma_0}\left(1+\frac{1}{2}\gamma_0^{ij}
  \gamma_{2\, ij} + \ldots \right) \nonumber \\
  &=& \sqrt{-{\rm det}\gamma_0} + \frac{1}{2} \sqrt{-{\rm det}
  \gamma_0} \gamma_0^{ij} \gamma_{2\, ij} + \ldots \, .
  \label{expand}
 \eq
 For the zero-order (in the fluctuations) piece we can write:
 \be\label{root_g_0}
  \sqrt{-{\rm det}\gamma_0}\, d^2\sigma={\rm det}(e_0)^\mu_i d^2\sigma
  =\frac{1}{2}\epsilon_{\mu\nu} e_0^{\; \mu} e_0^{\; \nu} \, ,
 \ee
 where our conventions are such that $\epsilon_{01}=1$.
 This can be therefore cancelled by choosing a closed $B_{\mu\nu}$
 field as in section \ref{non_rel_lim}.  We are left with the
 second-order piece, which corresponds to the non-relativistic
 string action (\ref{S_NR}) we obtained in the last section.
 Indeed, from equations (\ref{S_nambu}), (\ref{g_2}) and
 (\ref{expand}) we get:
 \be\label{S_2}
  S_2=-\frac{T}{2} \! \int \! a(x^0)^2 \sqrt{-{\rm det}\gamma_0}
  \gamma_0^{ij} \delta_{ab}\pd_i X^a \pd_j X^b d^2\sigma =
  -\frac{T}{2} \! \int \! \sqrt{-{\rm det}\gamma_0} \gamma_0^{ij}
  G_{ab}(x^0)\pd_i X^a \pd_j X^b d^2\sigma \, .
 \ee

 \section{\label{hamiltonian} Hamiltonian Formulation}

 Here we discuss the Hamiltonian formulation of the Nambu-Goto and
 non-relativistic strings.

  \subsection{\label{rel_ham} Relativistic String}

 We first consider the relativistic Nambu-Goto string.  The Hessian
 of the Lagrangian (\ref{nambu}) has two null eigenvalues and as a
 result the canonical variables $x^M$, $p_M=\frac{\pd{\cal L}_{NG}}
 {\pd\dot x^M}$ satisfy two primary constraints, namely
 \bq
  && \frac{p_M p_N}{T}G^{MN}(x^0) + T x^{\prime M}x^{\prime N}
  G_{MN}(x^0) = 0  \label{energy_constr_rel}  \\
  && p_M x^{\prime M} = 0 \, ,  \label{pxp_constr_rel}
 \eq
 which are first-class.  From these the Dirac Hamiltonian can be
 constructed
 \be\label{hamilt_rel}
  H = \! \int \! {\cal H} d \sigma = \! \int \! \left[
  \frac{f(\sigma,\tau)}{2} \left( \frac{p_M p_N}{T}G^{MN}(x^0)
  + T x^{\prime M} x^{\prime N} G_{MN}(x^0) \right) +
  h(\sigma,\tau) p_M x^{\prime M}\right] d \sigma \, ,
 \ee
 where the Lagrange multipliers $f,g$ are arbitrary functions on the
 worldsheet.  The Poisson Brackets for $x^M$ and $p_M$ are
 \bq
  && \{X^M(\sigma), P_N(\sigma^\prime)\} =
     \delta(\sigma-\sigma^\prime) \delta^M_N \label{XP_rel} \\
  && \{X^M(\sigma), X^N(\sigma^\prime)\} = 0 \label{XX_rel} \\
  && \{P_M(\sigma), P_N(\sigma^\prime)\} = 0 \label{PP_rel}
 \eq
 and, choosing $h=0$, the equations of motion read:
 \be\label{hamilt_eom_rel}
  \left\{
        \begin{array}{lll}
         \dot x^M &=& \{x^M,\, H\} = f T^{-1} p_N(\sigma) G^{NM}(x^0)  \\
         \dot p_M &=& \{p_M,\, H\} = T \left(f x^{\prime N}
         \right)^{\prime} G_{NM}(x^0) + \frac{\dot a}{a} f \left(
         T^{-1} p_N p_\Lambda G^{N\Lambda}(x^0) - T x^{\prime N}
         x^{\prime \Lambda} G_{N\Lambda}(x^0) \right) \delta_M^0
        \end{array}
   \right.
  \ee
 Thus, for the spacelike fields $x^I$, $I=1,\ldots,D$, we have:
 \bq\label{eom_x_rel}
 \ddot x^I &=& f T^{-1} \left(\dot p_N G^{NI}(x^0) + p_N
  \dot G^{NI}(x^0) \right) + \dot f T^{-1} p_N G^{NI}(x^0) \nonumber \\
  &=& f \left(f x^{\prime \Lambda} \right)^{\prime} \delta_\Lambda^I
  -2\frac{\dot a}{a} f T^{-1} p_J a(x^0)^{-2} \delta^{JI} + \dot f
  T^{-1} p_J a(x^0)^{-2} \delta^{JI} \, .
 \eq
 In the temporal gauge $\dot x^0=1$ we have from equation
 (\ref{hamilt_eom_rel}) that
 \be\label{temp_gauge_rel}
  f = - T a(x^0)^2 p_0^{-1}
 \ee
 and so
 \be\label{p_I}
  p_I=T a(x^0)^2 f^{-1} \dot x^I = - p_0 \dot x^I \, .
 \ee
 Then equation (\ref{eom_x_rel}) becomes
 \be\label{eom_x_rel1}
  \ddot x^I =  T^2 a(x^0)^4 p_0^{-1} \left(p_0^{-1} x^{\prime I}
  \right)^{\prime} - \dot p_0 p_0^{-1} \dot x^I  \, .
 \ee
 Now $p_0$ can be found from the constraint (\ref{energy_constr_rel}):
 \be\label{p_0_eps}
  p_0=T a(x^0)^2 \left( \frac{ {{x^{\prime I}}^2} } {1-\dot x^{I 2}}
  \right)^{1/2} \!\! = \; T a(x^0)^2 \epsilon  \, .
 \ee
 In view of our previous results for $\dot \epsilon$ in the Lagrangian
 formulation (see equation (\ref{eom_eps})) we see that we are going to
 recover the equation of motion for $x^I$.  Indeed, from
 (\ref{hamilt_eom_rel}) we have:
 \be\label{p_0_dot_rel}
  p_0^{-1} \dot p_0 = - \dot a a T p_0^{-2} \left[ T^{-1} a(x^0)^{-2}
  \left( p_0^2 \dot x^{I 2}-p_0^2 \right) -  T a(x^0)^2 {x^{\prime I}}^2
  \right] = 2 \frac{\dot a}{a} \left( 1 - \dot x^{I 2} \right) \, ,
 \ee
 where we have used (\ref{p_0_eps}).  Thus, (\ref{eom_x_rel1}) becomes:
 \be\label{eom_fin_rel}
  \ddot x^I = \epsilon^{-1} \left(\epsilon^{-1} x^{\prime I}
  \right)^{\prime} - 2\frac{\dot a}{a} \left( 1 - \dot x^{I 2}
  \right) \dot x^I  \, ,
 \ee
 as in (\ref{eom_x}).  Finally, the constraint
 (\ref{pxp_constr_rel}) in this gauge becomes $\dot x^I
 x^{\prime I}=0$, as before.

 \subsection{\label{nonrel_ham} Non-Relativistic String}

 We now turn to the Hamiltonian formulation of the non-relativistic
 string.  The Hamiltonian density can be constructed from the
 constraints (\ref{energy_constr}-\ref{pxp_constr}) introducing
 arbitrary functions $f(\sigma,\tau)$ and $h(\sigma,\tau)$ as
 Lagrange multipliers:
 \be\label{Hamilt_dens}
  {\cal H} = f \left\{ p_\mu \epsilon^{\mu\rho} \eta_{\rho\nu}
  x^{\prime\nu} + \frac{1}{2} \left(\frac{P_aP_b}{T_0}G^{ab}(x^0)
  + T_0 X^{\prime a}X^{\prime b} G_{ab}(x^0) \right) \right\} +
  h \left( p_\mu x^{\prime\mu} + P_a X^{\prime a} \right) \, .
 \ee
 The canonical variables satisfy the following Poisson
 Brackets:
 \bq
  && \{X^M(\sigma), P_N(\sigma^\prime)\} =
     \delta(\sigma-\sigma^\prime) \delta^M_N \label{XP} \\
  && \{X^M(\sigma), X^N(\sigma^\prime)\} = 0 \label{XX} \\
  && \{P_M(\sigma), P_N(\sigma^\prime)\} = 0 \label{PP} \, .
 \eq
 Then, choosing $h=0$ the transverse equations of motion are:
 \be\label{hamilt_eom_transv}
  \left\{
         \begin{array}{lll}
          \dot X^a &=& \{X^a,\, H\}=f T_0^{-1}P_b(\sigma)G^{ab}(x^0) \\
          \dot P_a &=& \{P_a,\, H\} = T_0 \left(f X^{\prime b}
          (\sigma)\right)^{\prime} G_{ab}(x^0)
         \end{array}
  \right.
 \ee
 while the longitudinal ones read:
 \be\label{hamilt_eom_long}
  \left\{
         \begin{array}{lll}
          \dot x^\mu &=& \{x^\mu,\, H\} = f \epsilon^{\mu\rho}
          \eta_{\rho\nu} x^{\prime\nu}(\sigma) \\
          \dot p_\mu &=& \{p_\mu,\, H\} = \left( f p_\kappa(\sigma)
          \right)^{\prime} \epsilon^{\kappa\rho} \eta_{\rho\mu} +
          \frac{\dot a}{a} f \left(\frac{P_aP_b}{T_0}G^{ab}(x^0) -
          T_0 X^{\prime a} X^{\prime b} G_{ab}(x^0) \right) \delta_\mu^0
         \end{array}
  \right.
 \ee
 Now choose the gauge (\ref{nonrel_gauge}) by setting $\dot x^0=1$ and
 $x^{\prime 1}=\lambda$.  Equations (\ref{hamilt_eom_long}) then imply
 \be\label{f}
  f=-\lambda^{-1}
 \ee
 and the transverse equations of motion (\ref{hamilt_eom_transv}) give
 \be\label{eom_hamilt}
  \ddot X^a = -(\lambda T_0)^{-1} \left(\dot P_b G^{ab}(x^0) + P_b
  \dot G^{ab}(x^0) \right) = \lambda^{-2} X^{\prime\prime a} -
  2\frac{\dot a}{a} \dot X^a
 \ee
 recovering equation (\ref{eom_gauge}).

 From equations (\ref{hamilt_eom_long}) we have:
 \bq\label{p_0_dot_hamilt}
  \dot p_0 &=& \lambda^{-1} \left[p_1^\prime - \frac{\dot a}{a}
  \left( \frac{P_aP_b}{T_0}G^{ab}(x^0) - T_0 X^{\prime a}X^{\prime b}
  G_{ab}(x^0) \right) \right] \\
  &=& - \lambda^{-2} (P_a X^{\prime a})^\prime + \lambda^{-1}
  \frac{\dot a}{a} \left( T_0 X^{\prime a} X^{\prime b} G_{ab}(x^0)
  - \frac{P_aP_b}{T_0} G^{ab}(x^0) \right) \, ,
 \eq
 using the constraint (\ref{pxp_constr}).  Since $p_0$ is given
 by equation (\ref{energy_constr}) and $P_a=-\lambda T_0 \dot X_a$
 (from (\ref{hamilt_eom_long})) we recover equation (\ref{p_0_dot}).

 \section{\label{NR_EM_tensor} Energy-Momentum tensor of
          Non-Relativistic Action}

 In order to obtain the energy-momentum tensor of the non-relativistic
 string, we vary the action (\ref{S_NR}) with respect to the
 background metric $G_{MN}$:
 \be\label{EM_Tensor}
  T^{MN}=\frac{-2}{\sqrt{-{\rm det}G_{MN}}} \frac{\delta S}{\delta
        G_{MN}}=\frac{-2}{\sqrt{-{\rm det}G_{MN}}}
        \left(
         \begin{array}{ll}
           \frac{\delta S}{\delta G_{\mu\nu}} &       0            \\
                    0              & \frac{\delta S}{\delta G_{ab}}
         \end{array}
        \right)
 \ee
 The transverse part reads
 \be\label{EM_trans}
  T^{ab} = \frac{T_0}{\sqrt{-{\rm det}G_{MN}}} \! \int \! d^2\sigma
                \sqrt{-{\rm det}g} g^{ij} \pd_i X^a \pd_j X^b
                \delta^{(D+1)}(y^M-x^M(\sigma^i))
 \ee
 whereas the longitudinal part is:
 \bq
  T^{\mu\nu} &=& \frac{T_0}{\sqrt{-{\rm det}G_{MN}}} \! \int \!
               d^2\sigma \sqrt{-{\rm det}g} \left[\frac{1}{2} g^{mn}
               \pd_m x^\mu \pd_n x^\nu g^{ij} \pd_i X^a \pd_j
               X^b G_{ab} \right. \nonumber \\
               && \left. - g^{im}\pd_i X^a \pd_m x^\mu
               g^{jn}\pd_j X^b \pd_n x^\nu G_{ab} \right]
               \delta^{(D+1)}(y^M-x^M(\sigma^i)) \label{EM_long}
 \eq

 In particular, the $00$ component in the gauge (\ref{nonrel_gauge})
 becomes
 \be\label{T00}
  T^{00}(\eta,y^K)=-\frac{1}{a(\eta)^{D+1}} \ \! \int \! \frac{\lambda T_0}{2}
  \left(\dot X^a\dot X^b + \lambda^{-2}X^{\prime a}X^{\prime b}\right)
  \delta_{ab} \delta^{(D)}(y^K-x^K(\sigma,\eta))  d\sigma \, ,
 \ee
 where, having integrated out $\delta(\eta-\tau)$, $K$ runs from 1 to
 $D$.  Note that the explicit scale-factor dependence of the integrand
 cancels, because of the presence of $\sqrt{-g}g^{-1}g^{-1}G$ in equation
 (\ref{EM_long}).  The first factor scales like $a(\eta)^2$, the next
 two factors as $a(\eta)^{-2}$ each, and the last factor as $a(\eta)^2$,
 giving a scale-factor-independent result.  Of course $T^{00}$ still
 depends on time through the time dependence of the fields $X^a$.

 The string energy can be defined as in section \ref{rel_string}, by
 considering a spacial hypersurface $\eta\!=\!{\rm const}$, with normal
 (co)vector $n_M\!=\!(a(\eta),{\bf 0})$, and integrating the energy
 density $n_M n_N T^{MN}=a(\eta)^2 T^{00}=T^0_{\;0}$ over the relevant
 D-volume:
 \bq\label{energy_final}
  E(\eta) &=& - \! \int \! \sqrt{h} n_M n_N T^{MN} d^D{\bf y} \nonumber \\
         &=& - \! \int \! a(\eta)^{D+2}  T^{00} d^D{\bf y} \nonumber \\
         &=& a(\eta) \frac{\lambda T_0}{2} \! \int \!  d\sigma \left(
             \dot X^a \dot X^b + \lambda^{-2} X^{\prime a} X^{\prime
             b} \right) \delta_{ab} \, .
 \eq

 Similarly, the $ab$ components of the energy-momentum tensor in the
 gauge (\ref{nonrel_gauge}) are
 \be\label{Taa}
  T^{ab}(\eta,y^K)=\frac{1}{a(\eta)^{D+1}} \ \! \int \! \lambda T_0
  \left(-\dot X^a\dot X^b + \lambda^{-2}X^{\prime a}X^{\prime b}\right)
  \delta^{(D)}(y^K-x^K(\sigma,\eta))  d\sigma \, .
 \ee

\bibliographystyle{JHEP.bst}
\bibliography{NRstring}

\end{document}